\documentclass[]{aastex631}

\usepackage{amsmath}
\usepackage[ruled,vlined]{algorithm2e}
\usepackage{rotating}

\DeclareMathOperator*{\argmax}{arg\,max}
\DeclareMathOperator*{\argmin}{arg\,min}

\shorttitle{Optimal scheduling of binary stars observations}
\shortauthors{Videla et al.}
\graphicspath{{./}{figures/}}
\begin{document}

\title{Optimal observational scheduling framework for binary and multiple stellar systems\footnote{Based in part on observations obtained at the international Gemini Observatory, a program of NSF’s NOIRLab, which is managed by the Association of Universities for Research in Astronomy (AURA) under a cooperative agreement with the National Science Foundation on behalf of the Gemini Observatory partnership: the National Science Foundation (United States), National Research Council (Canada), Agencia Nacional de Investigaci\'on y Desarrollo (Chile), Ministerio de Ciencia, Tecnolog\'{i}a e Innovaci\'on (Argentina), Minist\'erio da Ci\^{e}ncia, Tecnologia, Inova\c{c}\~{a}es e Comunica\c{c}\~{o}es (Brazil), and Korea Astronomy and Space Science Institute (Republic of Korea). Based also in part on observations obtained at the Southern Astrophysical Research (SOAR) telescope, which is a joint project of the Minist\'{e}rio da Ci\^{e}ncia, Tecnologia, e Inova\c{c}\~{a}oes (MCTI) da Rep\'{u}blica Federativa do Brasil, the U.S. National Optical Astronomy Observatory (NOAO), the University of North Carolina at Chapel Hill (UNC), and Michigan State University (MSU).}}

\correspondingauthor{Miguel Videla}
\email{miguel.videla@ug.uchile.cl}

\author[0000-0002-7140-0437]{Miguel Videla}
\affiliation{Department of Electrical Engineering\\
Information and Decision Systems Group (IDS) \\
Facultad de Ciencias Físicas y Matemáticas, Universidad de Chile \\
Beauchef 850, Santiago, Chile}

\author[0000-0003-1454-0596]{Rene A. Mendez}
\affiliation{Department of Astronomy \\
Facultad de Ciencias Físicas y Matemáticas, Universidad de Chile \\
Casilla 36-D, Santiago, Chile \\
and \\
European Southern Observatory \\
Alonso de Córdova 3107, Vitacura\\
Santiago, Chile}

\author[0000-0002-0256-282X]{Jorge F. Silva}
\affiliation{Department of Electrical Engineering\\
Information and Decision Systems Group (IDS) \\
Facultad de Ciencias Físicas y Matemáticas, Universidad de Chile \\
Beauchef 850, Santiago, Chile}

\author{Marcos E. Orchard}
\affiliation{Department of Electrical Engineering\\
Information and Decision Systems Group (IDS) \\
Facultad de Ciencias Físicas y Matemáticas, Universidad de Chile \\
Beauchef 850, Santiago, Chile}

%% Note that the \and command from previous versions of AASTeX is now
%% depreciated in this version as it is no longer necessary. AASTeX 
%% automatically takes care of all commas and "and"s between authors names.

%% AASTeX 6.31 has the new \collaboration and \nocollaboration commands to
%% provide the collaboration status of a group of authors. These commands 
%% can be used either before or after the list of corresponding authors. The
%% argument for \collaboration is the collaboration identifier. Authors are
%% encouraged to surround collaboration identifiers with ()s. The 
%% \nocollaboration command takes no argument and exists to indicate that
%% the nearby authors are not part of surrounding collaborations.

%% Mark off the abstract in the ``abstract'' environment. 
\begin{abstract}
The optimal instant of observation of astrophysical phenomena for objects that vary on human time-sales is an important problem, as it bears on the cost-effective use of usually scarce observational facilities. In this paper we address this problem for the case of tight visual binary systems through a Bayesian framework based on the maximum entropy sampling principle. 
%Our proposed information-driven methodology exploits the periodic nature of binary systems, allowing to characterize the observation uncertainty in a dense and meaningful range of time at a low computationally cost, and providing an estimation of the probability distribution of the optimal observation time.
Our proposed information-driven methodology exploits the periodic structure of binary systems to provide a computationally efficient estimation of the probability distribution of the optimal observation time.
We show the optimality of the proposed sampling methodology in the Bayes sense and its effectiveness through direct numerical experiments. We successfully apply our scheme to the study of two visual-spectroscopic binaries, and one purely astrometric triple hierarchical system. We note that our methodology can be applied to any time-evolving phenomena, a particularly interesting application in the era of dedicated surveys, where a definition of the cadence of observations can have a crucial impact on achieving the science goals.
\end{abstract}

%% Keywords should appear after the \end{abstract} command. 
%% The AAS Journals now uses Unified Astronomy Thesaurus concepts:
%% https://astrothesaurus.org
%% You will be asked to selected these concepts during the submission process
%% but this old "keyword" functionality is maintained in case authors want
%% to include these concepts in their preprints.

%% From the front matter, we move on to the body of the paper.
%% Sections are demarcated by \section and \subsection, respectively.
%% Observe the use of the LaTeX \label
%% command after the \subsection to give a symbolic KEY to the
%% subsection for cross-referencing in a \ref command.
%% You can use LaTeX's \ref and \label commands to keep track of
%% cross-references to sections, equations, tables, and figures.
%% That way, if you change the order of any elements, LaTeX will
%% automatically renumber them.
%%
%% We recommend that authors also use the natbib \citep
%% and \citet commands to identify citations.  The citations are
%% tied to the reference list via symbolic KEYs. The KEY corresponds
%% to the KEY in the \bibitem in the reference list below. 

\section{Introduction} \label{Sec:intro}

Technological advances in observational astronomy have allowed the discovery of a plethora of celestial objects of great interest to the astronomical community, motivating numerous observational campaigns to collect data for their study. Unfortunately, the number of celestial objects and physical phenomena exceeds by much our observational capacity due to the limited number of high-precision observational facilities. This restriction causes strong competition among astronomers for precious observational time, limiting the number of observations that can be made \citep{patat2012growth}\footnote{More recent examples of the over-subscription rates for some facilities to which our team has access can be found at \url{https://www.gemini.edu/observing/science-operations-statistics}, \url{https://noirlab.edu/science/observing-noirlab/observing-ctio/observing-soar/observing-statistics}, or \url{http://www.eso.org/sci/observing/phase1/p111/pressure.html}.}. It is therefore relevant, in particular for the study of those dynamical systems that evolve on human time-scales, to determine the most adequate instant(s) of observation(s) which allows to characterize the system under study as best as possible. In this context, the observation scheduling problem can naturally be addressed in the framework of a now mature theory (see below) that has been developed precisely for the optimal design of experiments.

In this paper we apply optimal experimental design theory for the study of binary and multiple stellar systems to define an observational planning that is the most informative regarding the physical parameters of these systems (e.g., orbital elements, component masses). In particular, our objects of interest are tight binaries, that require for their proper study the use of (scarce) specialized Rayleight-limited interferometric (e.g., Speckle imaging), or high-angular resolution (e.g., Adaptive Optics) techniques on large-aperture telescopes to be able to resolve them into individual components for a full determination of their orbital architecture. However, we emphasize that our methodology (described in Section~\ref{Sec:2}) can be naturally extended to any other time-evolving phenomena, which might be a particularly interesting application aspect in the era of big dedicated surveys, where a definition of the cadence of observations is a crucial aspect of the science goals (e.g., in the study of exo-planets, \citet{ford2008adaptive,loredo2012bayesian}).

The optimal experimental design framework aims to determine an experiment that is optimal according to some statistical criterion, allowing to optimize the costs involved in the estimation of statistical models \citep{pukelsheim2006optimal}. A common criterion for determining an optimal design is to minimize the variance of the estimators, which must be adapted according to the properties of the statistical model considered. In the case of multivariable linear statistical models, the variance is represented by the covariance matrix, and hence, the problem of minimizing the variance of the estimates restates to minimize invariant transformations of the Fisher information matrix (the inverse of the covariance matrix) according to some optimality criteria \citep{box1982choice}. Some commonly used optimality criteria consists on minimizing the trace of the covariance matrix (A-optimality), minimizing the variance of a best linear unbiased estimator of a predetermined linear combination of model parameters (C-optimality), maximizing the determinant of the information matrix (D-optimality), maximizing the eigenvalues of the information matrix (E-optimality), among others. These are known as alphabetical design criteria.

The alphabetical design criteria are initially restricted to linear models and do not consider prior information about the inference parameters, but these limitations can be relaxed. The Bayesian experimental design approach overcomes this last limitation by addressing the incorporation of a prior distribution of the inference parameters into the experimental design setting. In this new setting, the original problem restates to maximize the expected utility of an experiment according to some utility function that reflects the purpose of the experiment \citep{lindley1972bayesian,chaloner1995bayesian}. Some Bayesian equivalents of the previously mentioned alphabetical design criteria can be recovered by choosing an appropriate utility function in some specific contexts; e.g., by considering the Shannon utility function in Gaussian linear models, the Bayesian D-optimality criteria can be derived \citep{bernardo1979expected}. In the nonlinear case, the determination of the expected utility of an experiment often involves the computation of complicated integrals, usually requiring approximations of some elements of the problem, e.g., considering a normal approximation to the posterior distribution \citep[pp.~224]{james1985statistical}, approximating the prior distribution by a one-point distribution through the local optimality approach \citep{chernoff1953locally}, or estimating the expected utility through Markov chain Monte Carlo methods \citep{muller1996numerical}.

Although the optimal experimental design problem has been vastly studied and applied in diverse contexts, such as in biology \citep{balsa2008computational}, psychology \citep{myung2009optimal}, or material science \citep{dehghannasiri2017optimal}, its application in the astronomical context has not been so abundant. In this last context, the optimal experimental design works has been mainly focused on determining the optimal instant of observations for certain celestial phenomena. The earliest works on this field were limited to non-adaptive schedules (i.e., those independent from previous observations), such as regular periodic or random uniform schedules \citep{sozzetti2002narrow,ford2004choice}. Later, statistical rigorous approaches based on the Bayesian optimal experimental design theory were also explored. \cite{ford2008adaptive} and \cite{loredo2012bayesian} proposed adaptive schedules approaches for the search and the characterization of the radial velocity of exoplanets, respectively. Both approaches were based on the maximum entropy sampling principle \citep{shewry1987maximum}, which aims to maximize the information about the studied object by observing the zones of highest predicted uncertainty. Recently, a different adaptive scheduling approach, based on the Fisher information matrix, was proposed \citep{hees2019adaptive}. Even though this last approach presents a superior computational efficiency, it was restricted only to Gaussian problems.

In the case of binary systems, the commonly adopted criterion to perform observations is simply to ensure adequate orbital coverage, however defining exactly what is adequate has not ---to the best of our knowledge--- been discussed in a mathematically rigorous manner. For example, earlier work by \citet{lucy2014mass} has shown that, in order to obtain a reliable orbit, the phase coverage has to include at least about 40\% of the orbit, but there is no indication on that work where in the orbit (when) these observations should be placed (or made). The present work adapts the optimal observation scheduling approach based on the maximum entropy sampling criterion, previously developed by \cite{ford2008adaptive} and \cite{loredo2012bayesian} for exoplanets research, to the study of binary and hierarchical stellar systems. The proposed method uses the estimated posterior distribution of binary stellar systems to generate a probability distribution of the observation time that provides the highest information gain about the system´s parameters. Remarkably, this information-driven approach allows characterizing the posterior predictive information gain in a dense and meaningful range of time with very modest computational costs, exploiting the periodic nature of binary systems to bound the range of times to evaluate, and providing an estimation of the probability distribution of the optimal observation time. The software is freely available on GitHub\footnote{\texttt{BinaryStars} codebase: \url{https://github.com/mvidela31/BinaryStars}.} under a 3-Clause BSD License.

The paper is organized as follows: In Section~\ref{Sec:2}, we introduce the Bayesian experimental design problem and the maximum entropy sampling criterion. In Section~\ref{Sec:3}, we introduce the orbital statistical model and our proposed optimal scheduling methodology for astrometric observations of binary stars. In Section~\ref{Sec:4}, we evaluate the proposed methodology on three different binary stellar systems (including a hierarchical triple star) as tests cases of our approach, thoroughly analyzing and discussing the obtained results. Finally, in Section~\ref{Sec:conclu} we provide the main conclusions and outlook of our work.

\section{Optimal experimental design \& maximum entropy Sampling}
\label{Sec:2}

The current section introduces a formalization for the Bayesian experimental design problem in terms of a maximum entropy sampling criterion; providing insights on the equivalence of these problems under certain specific conditions.

\subsection{Bayesian experimental design}
\label{Sec:bed}

The Bayesian experimental design \citep{lindley1972bayesian} is a decision-theoretic approach to the experimental design problem that allows to incorporate prior knowledge on the parameters to be estimated, as well as the uncertainty of the observations.

Let $\theta\in \Theta$ be the vector of parameters that we want to estimate from a random vector of observations $Y\in\mathcal{Y}$, and let $\xi\in\Xi$ be an experimental design to be chosen. Based on the observations $Y$, a decision $d\in D$ will be made. The decision procedure consist of first selecting an experiment design $\xi$, and then choosing a terminal decision $d$.

Thereby, the expected utility of the best decision $d\in D$ for any experiment $\xi\in\Xi$ is given by:
\begin{equation}\label{eq_utility}
    U(\xi) \equiv\int_Y \max_{d\in D}\int_\Theta p(y|\xi) \cdot p(\theta|y,\xi) \cdot U(d,y,\theta,\xi) \, d\theta \, dy,
\end{equation}
where $p(y|\xi)=\int p(\theta)p(y|\theta,\xi)d\theta$ represents the marginal distribution in the observation space for the experiment $\xi$, $p(y|\theta,\xi)$ denotes the likelihood of $Y$ given $\theta$ for the experiment $\xi$, $p(\theta|y,\xi)$ denotes the posterior distribution of $\theta$ for the experiment $\xi$, $p(\theta)$ is the prior distribution of the parameters (which is independent of $\xi$), and $U(d,y,\theta,\xi)$ represents a general utility function. The utility function is a real-valued function that describes the worth (based on the experimental purpose) of making a decision $d\in D$ about choosing an experiment $\xi\in\Xi$ for the observations $y\in\mathcal{Y}$ and the model, indexed by $\theta\in\Theta$. $U(\xi)$ can be seen as the average reward we achieve in choosing $\xi$, for the task of estimating $\theta$ from $Y$. 

In this context, the Bayesian experimental design can be defined as a procedure that aims to determine the experiment $\xi^*\in\Xi$ that maximizes the
expected utility of the best decision:
\begin{equation}
    \xi^* \equiv \argmax_{\xi\in\Xi} U(\xi).
\label{eq:opt_design}
\end{equation}

An extensive list of expected utility functions for different design criteria can be found in the review of \citet{chaloner1995bayesian}.

\subsection{Maximum information extraction}
\label{Sec:mie}
A standard way to define the expected utility of the best decision $U(\xi)$ for Equation~(\ref{eq:opt_design}) is to consider it as the amount of information that an experiment $\xi\in\Xi$ offers on the unknown vector of parameters $\Theta$, given the acquired observations of $Y$. More precisely, the concept of information can be defined in terms of the {\em Shannon entropy} \citep{cover1999elements}, which measures the uncertainty (prior) or average information that a random variable has, after measuring it. Let $X$ be a discrete random variable with alphabet $\mathcal{X}$ and probability mass function $p(x)=\mathbb{P}(X=x), x\in\mathcal{X}$, the entropy of $X$ is defined as follows:
\begin{equation}
    H(X)=\sum_{x\in\mathcal{X}} -p(x) \cdot \log p(x).
\end{equation}
In the case of continuous random variables, the concept of differential entropy is used instead \citep{cover1999elements}. Let $X$ a continuous random variable with support on $\mathbb{X}$ and probability density function $f(x)$, then the differential entropy of $X$ is defined as follows:
\begin{equation}
    h(X)=\mathbb{E}_{X\sim f(x)}\{-\log f(X)\}=-\int_{\mathbb{X}} f(x) \cdot \log f(x) \, dx.
\end{equation}
where $\mathbb{E}\{.\}$ denotes the expectation of the argument.

Thereby, the {\em Shannon utility} $U_S(\xi)$ 
is the information gain or, equivalently, the uncertainty reduction (in the Shannon sense) from the prior distribution $(p(\theta))_{\theta\in \Theta}$ to the posterior distributions $(p(\theta|y,\xi))_{(\theta,y)\in \Theta \times \mathcal{Y}}$ \citep{lindley1956measure}. For this objective, the Shannon utility function $U_S(y,\theta,\xi)$ is defined in terms of the log-likelihood ratio between the mentioned distributions:
\begin{equation}\label{eq_likelihood_ratio}
U_S(y,\theta,\xi)= \log(p(\theta|y,\xi)) - \log(p(\theta))=\log \frac{p(\theta|y,\xi)}{p(\theta)},
\end{equation}
and hence, the average utility of the experiment $\xi$ in Equation~(\ref{eq_utility}) can be written as:
\begin{align}\label{eq_shannon_u}
    U_S(\xi)&=\int_Y \int_\Theta p(y|\xi) \cdot p(\theta|y,\xi) \cdot \log \frac{p(\theta|y,\xi)}{p(\theta)} \, d\theta \, dy  \\
    &=\int_Y \int_\Theta p(y|\xi) \cdot p(\theta|y,\xi) \cdot \log(p(\theta|y,\xi)) \, d\theta \, dy - \int_Y \int_\Theta \underbrace{p(y|\xi) \cdot p(\theta|y,\xi)}_{{p(\theta,y|\xi)=p(\theta)p(y|\theta,\xi)}} \cdot \log(p(\theta)) \, d\theta \, dy  \\
    &=\int_Y p(y|\xi)\int_\Theta p(\theta|y,\xi) \cdot \log(p(\theta|y,\xi)) \, d\theta \, dy - \int_\Theta p(\theta) \cdot \log(p(\theta)) \int_Y p(y|\theta,\xi) \, dy \, d\theta  \\
    &=-\mathbb{E}_{Y\sim p(y|\xi)}\{H(\Theta|Y,\xi)\} + H(\Theta)  \label{eq_shannon_sum}\\
    \label{eq_shannon_u_c}
    &= I(\Theta; Y|\xi).
\end{align}

Indeed, the final expression in Equation~(\ref{eq_shannon_u_c}) is the celebrated {\em mutual information} between $Y$ and $\Theta$ for the experiment $\xi$ denoted by $I(\Theta; Y|\xi)$ \citep{cover1999elements}.
Consequently, the optimal Bayesian experimental design in Equation~(\ref{eq:opt_design}) can be restated as:
\begin{equation}
    \xi^*=\argmax_{\xi\in\Xi} I(\Theta; Y|\xi) 
\label{eq:opt_design_infogain}
\end{equation}
Since $H(\Theta)$ does not depend on the experiment $\xi$, the argument of the optimization problem in Equation~(\ref{eq:opt_design_infogain}) can be simply interpreted as the minimization (note the $-$ sign in Equation~(\ref{eq_shannon_sum})) of the expected posterior entropy $\mathbb{E}_{Y\sim p(y|\xi)}\{H(\Theta|Y,\xi)\}$ in Eq.(\ref{eq_shannon_sum}).
%conditional entropy $ H(\Theta|Y,\xi) = \mathbb{E}_{\Theta,Y\sim p(\theta) p(y|\theta,\xi) } (- \log(p(\Theta|Y,\xi)))$.

%From the expression in (\ref{eq_likelihood_ratio}), i
It is noteworthy that an alternative characterization for the utility $U_S(\xi)$ defined in Equation~(\ref{eq_likelihood_ratio}) may be considered: Indeed, from Equation~(\ref{eq_shannon_u})) this function can be regarded as the average {\em Kullback-Leibler divergence} \citep{cover1999elements} between the prior $(p(\theta))_{\theta\in \Theta}$ and posterior distributions $(p(\theta|y,\xi))_{(\theta,y)\in \Theta \times \mathcal{Y}}$: 
\begin{equation}
U_S(\xi)=\mathbb{E}_{Y\sim p(y|\xi)} \left\{ D_{KL}(p(\theta|Y,\xi)||p(\theta)) \right\} 
\end{equation}
%$U(y,\theta,\xi)=D_{KL}(p(\theta|y,\xi)||p(\theta))$.
where for all $y\in \mathcal{Y}$ and for all $\xi \in \Xi$ the Kullback-Leibler divergence \citep{cover1999elements} is defined as:
\begin{equation}
    D_{KL}(p(\theta|y,\xi)||p(\theta))=  \int_{\Theta} \log \frac{p(\theta|y,\xi)}{p(\theta)} \cdot p(\theta|y,\xi) \, d\theta.
\end{equation}

In the following section, we introduce the maximum entropy sampling criterion and its relationship with the optimal Bayesian experimental design problem just described. 

\subsection{Maximum entropy sampling}
\label{Sec:mes}
The maximum entropy sampling \citep{shewry1987maximum} is an information theoretic-based problem that proposes to select an experiment $\xi$, in this case the experiment refers to the sample, that maximizes the gain in information from predictions at unsampled sites.

In the information-theoretic language, the Maximum Entropy Sampling criterion attempts to choose a sample that minimizes the uncertainty or entropy of a system. Let $\Gamma_{\mathcal{S}}= (\Gamma_i)_{i\in \mathcal{S}}$ with $\mathcal{S}= \left\{{1,...,N}\right\}$ be a finite system described by a random vector $\Gamma_i$, with $i\in \mathcal{S}$ the index of an observation site. Let $s,\bar{s}$ be a binary partition of $\mathcal{S}$, i.e, $\mathcal{S}=s\dot{\cup}\bar{s}$, where $\Gamma_{s}=(\Gamma_i)_{i\in s}$ represents the observations or measurements taken to estimate the whole system $\Gamma_{\mathcal{S}}=(\Gamma_i)_{i\in \mathcal{S}}$. Therefore, the objective of the problem is 
%to interpolate the unobserved sites of the system $(\Gamma_i)_{i\in\bar{s}}$ from the observed portion of the system $(\Gamma_i)_{i\in s}$. 
to determine the sample set $s\subseteq\mathcal{S}$ that minimizes the average uncertainty of the unobserved portion of the system $\Gamma_{\bar{s}}$ given the observations of the system $\Gamma_s$, i.e., to minimize 
%...........
\begin{equation}
\mathbb{E}_{\Gamma_s}\{H(\Gamma_{\bar{s}}|\Gamma_s)\}
\label{eq:mes_or}
\end{equation}

By the chain rule property of the entropy \citep{cover1999elements} on $\Gamma_S$, the following expression can be obtained:
\begin{equation}
    H(\Gamma_{\mathcal{S}})=H(\Gamma_s)+\mathbb{E}_{\Gamma_s}\{H(\Gamma_{\bar{s}}|\Gamma_s)\}.
    \label{eq:mes}
\end{equation}
%Using this identity, the maximisation of the information on the system $\Gamma_{\mathcal{S}}$ given the samples $\Gamma_s$ to be acquired, is achieved by minimizing the term $\mathbb{E}_{\Gamma_s}\{H(\Gamma_{\bar{s}}|\Gamma_s)\}$ in Equation~(\ref{eq:mes}). Alternatively, 
Using this identity, an considering that $H(\Gamma_{\mathcal{S}})$ is fixed (i.e., independent of $s$) and finite, the aforementioned minimization of (\ref{eq:mes_or}) reduces to the maximization of $H(\Gamma_s)$; i.e., to select the samples
%.......
$$s^*=\argmax_{s\subseteq S} H(\Gamma_s)$$ 
with the highest observed uncertainty. In other words, the maximum entropy sampling criterion aims at selecting the observations with the maximum amount of uncertainty. Ensuring that objective, the unsampled population $\Gamma_{\bar{s}}$ exhibits minimum conditional uncertainty given the sampled population $\Gamma_{{s}}$. 

We must emphasize that this sampling criterion is not equivalent to the Bayesian optimal experimental design in Equation~(\ref{eq:opt_design_infogain}) since the figure of merit of this last problem is the minimization of the expected posterior entropy. Importantly, \cite{sebastiani2000maximum} have shown that both problems are equivalent under certain specific assumptions. Following the Bayesian experimental framework introduced in the previous section, let $y\in Y$ be the observations of a system, $\theta\in\Theta$ its parameters and $\xi\in\Xi$ an experiment. By choosing $\Gamma_s= Y|\xi$ and $\Gamma_{\bar{s}}= \Theta$ in Equation~(\ref{eq:mes}), the following expression is obtained:
\begin{equation}
    H(Y,\Theta|\xi)=H(Y|\xi)+\mathbb{E}_{Y\sim p(y|\xi)}\{H(\Theta|Y,\xi)\}.
    \label{eq:bayes_mes}
\end{equation}
If $H(Y,\Theta|\xi)$ is fixed in the sense that is independent of any experiment $\xi\in\Xi$, and all the terms in Equation~(\ref{eq:bayes_mes}) are finite, the minimization of the expected posterior entropy $\mathbb{E}_{Y\sim p(y|\xi)}\{H(\theta|Y,\xi)\}$ is achieved by maximizing the marginal entropy in the observation space $H(Y|\xi)$. In other words, the maximum entropy sampling criterion is equivalent to the Bayesian optimal experimental design problem if the joint Shannon entropy of $(Y,\Theta)$ is not functionally dependent on the experiment $\xi$ itself. This last condition is however generally not straightforward to guarantee.

Fortunately, the independence condition between the joint distribution of $(Y,\Theta)$ and the experiment $\xi$ can be reformulated. By interchanging the role of $Y$ and $\Theta$ in Equation~(\ref{eq:bayes_mes}), we get that:
\begin{equation}
    H(Y,\Theta|\xi)=H(\Theta)+\mathbb{E}_{\Theta\sim p(\theta)}\{H(Y|\Theta,\xi)\}.
    \label{eq:bayes_mes_interchanged}
\end{equation}
As $p(\theta)$ is experiment independent (see Section~\ref{Sec:bed}), the previously stated invariant condition of $H(Y,\Theta|\xi)$ with respect to $\xi$ is equivalent to ensure that the term $\mathbb{E}_\Theta\{H(Y|\theta,\xi)\}$ is independent from the experiment $\xi$. Importantly in our optimal scheduling methodology of the observations in binary stellar systems, this reformulated condition is easier to check:  
%than the one derived from Equation~(\ref{eq:bayes_mes}) to verify 
i.e., the equivalence between the maximum entropy sampling criterion and the Bayesian optimal experiment design problem. 
%This last independence condition will be particularly useful .

%%%%%%%%%%%%%%%%%%%%%%%%%%%%%%%%%%%%%%%%%%%%%%%%%%%%%%%%%
%%%%%%%%%%%%%%%%%%%%%%%%%%%%%%%%%%%%%%%%%%%%%%%%%%%%%%%%%
\section{Optimal scheduling of binary stars observations}
\label{Sec:3}

This section introduces the statistical model for binary stellar systems and proposes a novel optimal scheduling methodology for the observations in binary stellar systems based on the maximum entropy sampling described in Section~\ref{Sec:mes}.

\subsection{Statistical model}
\label{Sec:3.1}
The statistical model of binary stars presented in this section follows the one presented in \cite{videla2022bayesian}, but with minor notation changes that facilitate the formulation of the optimal scheduling methodology presented later in Section~\ref{Sec:3.2}.

Firstly, let us assume that the individual positional and radial velocity observations follow a Gaussian distribution, and let us consider uniform priors for model parameters (unless another specific prior distribution is considered). Let $\{t_i,X_i,Y_i\}_{i=1}^{n}$ be a set of $n$ positional observations of the companion star relative to the primary binary stellar system in rectangular coordinates. In addition, let $\{\bar{t}_i,V_{1i}\}_{i=1}^{n_1}$ and $\{\tilde{t}_i,V_{2i}\}_{i=1}^{n_2}$ be a set of $n_1$ and $n_2$ RV observations of the primary and companion stars, respectively. Given a vector of orbital parameters $\theta$ (fixed but unknown), each observation (measurement) distributes as an independent Gaussian distribution centered in the value obtained by the Keplerian model $f_{\text{kep}}^{*}(\theta,t)$ at a given epoch $t$, with a standard deviation equal to the corresponding observational error $\sigma_i$:
\begin{equation}\label{eq_obs_model}
    X_i\sim \mathcal{N}(f_{\text{kep}}^{x}(\theta,t_i),\sigma_i^2),\,\,
    Y_i\sim \mathcal{N}(f_{\text{kep}}^{y}(\theta,t_i),\sigma_i^2),\,\,
    V_{1i}\sim \mathcal{N}(f_{\text{kep}}^{v_1}(\theta,\bar{t}_i),\sigma_i^2),\,\,
    V_{2i}\sim \mathcal{N}(f_{\text{kep}}^{v_2}(\theta,\tilde{t}_i),\sigma_i^2),
\end{equation}
where $f_{\text{kep}}^{x}(\theta,t)$ and $f_{\text{kep}}^{y}(\theta,t)$ follow Equation~(\ref{eq:pos}), and $f_{\text{kep}}^{v_1}(\theta,t)$ and $f_{\text{kep}}^{v_2}(\theta,t)$ follow Equations~(\ref{eq:V_p}) and (\ref{eq:V_c}), respectively.\footnote{See Appendix \ref{App:orbital_models}, for a detailed presentation of the Keplerian model.}

Denoting the complete observation vector by $\mathcal{D}=\{t_i,X_i,Y_i\}_{i=1}^{n}\cup\{\bar{t}_i,V_{1i}\}_{i=1}^{n_1}\cup\{\tilde{t}_i,V_{2i}\}_{i=1}^{n_2}$, the log-likelihood of $\mathcal{D}$ given the vector of parameters $\theta$ is given by:
\begin{equation}
\begin{split}
    \log p(\mathcal{D}|\theta)=&\sum_{i=1}^{n}\log \mathcal{N}(X_i|f_{\text{kep}}^{x}(\theta,t_i),\sigma_i^2)+\sum_{i=1}^{n}\log \mathcal{N}(Y_i|f_{\text{kep}}^{y}(\theta,t_i),\sigma_i^2)\\&+\sum_{i=1}^{n_1}\log \mathcal{N}(V_{1i}|f_{\text{kep}}^{v_1}(\theta,\bar{t}_i),\sigma_i^2)+\sum_{i=1}^{n_2}\log \mathcal{N}(V_{2i}|f_{\text{kep}}^{v_2}(\theta,\tilde{t}_i),\sigma_i^2).
\end{split}
\label{eq:loglikelihood}
\end{equation}
The prior distribution of each scalar orbital parameter $\theta_i$ is modeled as a uniform distribution  on their valid physical range denoted by $\Theta_i\subset \mathbb{R}$ \citep[Appendix~A]{videla2022bayesian}. Therefore, the prior distribution (under the assumption of independence between components) of the orbital parameter vector is:
\begin{equation}
    \log p(\theta)=\sum_{i=1}^{|\theta|}\log U(\min \Theta_i,\max\Theta_i),
\end{equation}
with $\theta_i\in\Theta_i,\forall i \in \{1,...,|\theta|\}$, and $U$ is the probability density function of a uniform distribution between $\min \Theta_i$, and $\max \Theta_i$.

At the inference stage, according to the Bayes theorem, the posterior distribution, given the evidence $\mathcal{D}$, is proportional to the likelihood times the prior, i.e., $p(\theta|\mathcal{D})\propto p(\mathcal{D}|\theta)p(\theta)$. Therefore, the complete posterior distribution can be obtained through the implementation of any sampling technique \citep{hastings1970monte,geman1984stochastic,goodman2010ensemble,neal2011mcmc,hoffman2014no}.

For the specific cases of study in this paper, the double-line spectroscopic binary model is characterized by the vector of orbital parameters $\theta_{SB2}=\{P,T,e,a,\omega,\Omega,i,V_0,\varpi,q\}$, while the single-line spectroscopic binary model is represented by $\theta _{SB1}=\{P,T,e,a,\omega,\Omega,i,V_0,f/\varpi\}$. The visual-triple hierarchical stellar system is, in turn, characterized by the vector of orbital parameters $\theta _{triple}=\{P_{A_aA_b},T_{A_aA_b},e_{A_aA_b},a_{A_aA_b},\omega_{A_aA_b},\Omega_{A_aA_b},i_{A_aA_b}\}\cup\{q_{A_aA_b}\}\cup\{P_{AB},T_{AB},e_{AB},a_{AB},\omega_{AB},\Omega_{AB},i_{AB}\}$. For further details on the Keplerian orbital models, please see Appendix \ref{App:orbital_models} and \citet{videla2022bayesian}.

\subsection{Optimal observations scheduling}
\label{Sec:3.2}

Following the Bayesian framework, let $y\in Y$ and $\theta\in\Theta$ be the observations and parameters of a system, respectively. The maximum information extraction problem in Equation~(\ref{eq:opt_design_infogain}) is to find the experiment $\xi\in\Xi$ that minimizes the expected posterior entropy $\mathbb{E}_Y\{H(\Theta|Y,\xi)\}$. This framework can be used for determining the optimal observation time in binary stellar systems. 
Considering that $Y$ is the observation space (composed by the position and radial velocity space), $\Theta$ is the orbital parameters space (that characterize the stellar system), and $\xi$ is the instant (epoch) of observation of the system (denoted as $t$ hereinafter), the problem in Equation~(\ref{eq:opt_design_infogain}) is to find the observation epoch $t^*\in\mathcal{T}$ that maximizes the information about the orbital parameters of the system. For that purpose, we propose to use the maximum entropy sampling criterion introduced in Section~\ref{Sec:mes}.
%, or equivalently, that minimizes the posterior entropy $\mathbb{E}_Y\{H(\Theta|Y,t)\}$.

According to the Bayesian modeling of binary stellar systems introduced in Section~\ref{Sec:3.1}, it is assumed that each observation distributes as a Gaussian with mean determined by the Keplerian orbital model at time $t$, and variance determined by the corresponding observation errors. Therefore, let $Y=(Y_1,...,Y_n)$ be a finite set of possible new observations (positional or RV) of the stellar system, the statistical model satisfies that:
\begin{equation}
    Y_i|(\theta,t_i)=f_{\text{kep}}^{*}(\theta,t_i)+\epsilon_i,
\end{equation}
where $\epsilon_i\sim\mathcal{N}(0,\sigma_i^2)$, $\sigma_i$ is the observation error and $f_{\text{kep}}^{*}(\cdot)$ is the function that maps the times $t_i$ into the observation space $Y_i$ given the stellar system orbital parameters $\theta$.

We note that $Y_i|(\theta,t_i)$ is a regression model according to \cite{sebastiani2000maximum}, and hence, if we additionally assume that the random variables $\epsilon_i$ is independent of $f_{\text{kep}}^{*}(\theta,t_i)$ and $\sigma_i=\sigma,\forall i\in  \left\{1,...,n\right\}$, then, we have that $H(Y|\theta,t)=H(\epsilon)$, i.e., $H(Y|\theta,t)$ is functionally independent from the choice of the observation time $t$ (the experiment). The fulfillment of this condition is sufficient to prove the equivalence between the maximum information extraction problem presented in Section~\ref{Sec:mie}, and the maximum entropy sampling criterion described in Section~\ref{Sec:mes}.

It is worth mentioning that assuming that $\sigma_i=\sigma$ for any $i\in  \left\{1,...,n\right\}$ implies that all the possible future observations must have the same uncertainty. In observational astronomy, this condition implies that no matter the value of the future observation (position in the coordinate space or magnitude of the radial velocity) the instrument (the telescope) measures the observational evidence with the same precision. Of course, this assumption is valid/realistic if future observations are acquired with the same instrumental setup, but it is violated when instruments of different precision are considered.

Therefore, using Equation~(\ref{eq:bayes_mes}) for the available observations $\mathcal{D}$ (i.e., $\Gamma_s\equiv Y|\mathcal{D},t$ in Equation~(\ref{eq:mes})), and noting that the discussed assumption makes the term $H(Y,\Theta|\mathcal{D},t)$ functionally independent of the value of $t$, then the observation time that maximizes the information about the orbital parameters of the system, i.e., the time that minimizes the expected posterior entropy, is the instant $t^*\in \mathcal{T}$ that maximizes the marginal entropy of the observations space:
\begin{equation}
    t^*=\argmin_{t\in \mathcal{T}}\mathbb{E}_{Y\sim p(y|t)}\{H(\Theta|Y,\mathcal{D},t)\}=\argmax_{t\in \mathcal{T}} H(Y|\mathcal{D},t).
    \label{eq:opt_obs_time}
\end{equation}

\subsection{The normalized scheduling}
\label{Sec:3.3}
We observe that the scheduling problem in Equation~(\ref{eq:opt_obs_time}) has a practical problem. The decision space $\mathcal{T}$ is unbounded, and there is no simple analytical way to tackle this task.
%, where we could choose a fixed range of $\mathcal{T}$ to make the problem in Equation~(\ref{eq:opt_obs_time}) tractable, but we would not be certain of the significance of this choice for the uncertainty representativeness of the derived entropy analysis for the studied object. In other words, if we choose $\mathcal{T}=[t_0,t_0+\delta]$, with $t_0,\delta\geq 0$, we would not know which value of $\delta$ makes the uncertainty analysis significant.]
To address this issue, we study the intrinsic regularity of the problem. Indeed, we notice that each selection of orbital parameter $\theta\in\Theta$ induces an orbit in the observation space with a given orbital period $P$. This specific period generates a periodic pattern on the information that the observation at time $t$ offers (a $P$-periodic information pattern). From this recurring pattern, we can reduce the range of exploration of $t$ to a bounded normalized domain, which makes the problem in Equation~(\ref{eq:opt_obs_time}) tractable.   
%a suitable selection of the set of observation times $\mathcal{T}$ must be defined to ensure a full exploration of each system $\theta$.

In summary to overcome the implementation challenge described above, we propose to replace the observation time random variable in Equation~(\ref{eq:opt_obs_time}) (the experiment) by the normalized observation time $\tau=(t-t_0)/P$ (the system's phase), where $P$ is the period of each orbital parameters vector $\theta\in\Theta$ and $t_0$ is the initial time of observation (a fixed design variable). Thereby, we only need to compute the marginal observation entropy $H(Y|\mathcal{D},\tau)$ on the closed interval $\tau\in[0,1]$ to evaluate the uncertainty of the full-completion posterior orbits, providing thus a complete exploration of each system $\theta\in\Theta$. Therefore, the optimal observation time problem in Equation~(\ref{eq:opt_obs_time}) reduces to:
\begin{equation}
    \tau^*=\argmin_{\tau\in [0,1]} \mathbb{E}_{Y\sim p(y|\tau)}\{H(\Theta|Y,\mathcal{D},\tau)\}=\argmax_{\tau\in [0,1]} H(Y|\mathcal{D},\tau).
    \label{eq:opt_obs_tau}
\end{equation}
From the normalized solution $\tau^*$, we can return to the un-normalized domain using the random variable $P$. This induces a probability distribution for the time of optimal observation $t^*$:
\begin{equation}
    t^*=\tau^*P+t_0.
\end{equation}
This reformulation of the original problem, Equation~(\ref{eq:opt_obs_time}) allows us to provide not only a deterministic estimate of the optimal epoch of observation $t^*$, but a complete distribution of it, with all its uncertainty being encapsulated by the temporal random variable corresponding to the system's period $P$.

Note that the normalized time $\tau\in[0,1]$ in Equation~(\ref{eq:opt_obs_tau}) can be interpreted as the percentage of a period completion (in decimal notation) from a starting time $t_0$ of all the orbits generated by the projection of the posterior distribution $p(\Theta|\mathcal{D})$ in the observation space. For example, when $\tau=0.5$ we estimate the entropy of the points of the half-complete orbits, while when $\tau=1$ we estimate the entropy of the points of the complete orbit. This approach differs from the original problem in Equation~(\ref{eq:opt_obs_time}) since it estimates the entropy of the projection of the orbits in a determined observation time $t\in\mathcal{T}$, completely ignoring the inherent periodicity of these astronomical objects. 

\subsection{Algorithmic implementation}
\label{Sec:3.4}
On the implementation, our optimal scheduling methodology involves solving Equation~(\ref{eq:opt_obs_tau}). This requires estimating the entropy of the posterior predictive distribution $p(y|\mathcal{D},t)=\int_{\Theta}p(y|\theta,\mathcal{D},t)p(\theta|\mathcal{D},t)d\theta$ at given normalized observation times $\tau=\frac{(t-t_0)}{P}\in[0,1]$. Fortunately, the samples of the posterior predictive distribution can be directly obtained by projecting samples of the posterior distribution into the observation space (as was made by, e.g., \citet{videla2022bayesian}), while the samples of the posterior distribution can be estimated using any Markov Chain Monte Carlo method (e.g., in \citet{gregory2005bayesian,gregory2011bayesian,hou2012affine,nelson2013run,mendez2017orbits,videla2022bayesian}). 
Finally, the estimation of the marginal posterior entropy can be computed using any entropy estimation algorithm (e.g., \cite{parzen1962estimation,kozachenko1987sample}). The steps involved in our optimal scheduling methodology are summarized in Algorithm \ref{alg:optimal_scheduling}.
\begin{algorithm}[H]
  \SetKwInOut{Parameter}{Parameters}
  \Parameter{$n,m,t_0,p(\theta|\mathcal{D})$}
  Initialise $\mathbf{D}_{n,m}$, $\mathbf{h}_n $, $\mathbf{p}_n$, $\mathbf{t}_n^*$\\
  \For{$0 \leq i < n$}{
    Sample from the posterior distribution\\
    $\mathbf{\theta} \sim p(\theta|\mathcal{D})$\\
    \SetKwFunction{getPeriod}{\textsc{getPeriod}}
    Get the orbital period\\
    $\mathbf{p}_i=$\getPeriod{$\mathbf{\theta}$}\\
    Project samples on the observation space\\
    \For{$0 \leq j < m$}{
        $\tau=j / (m - 1)$\\
        $t = \tau \cdot \mathbf{p}_i + t_0$\\
        $\mathbf{D}_{i,j}=f_{\text{kep}}^{*}(\mathbf{\theta},t)$
    }
  }
  Estimate the marginal observation entropy\\
  \For{$0 \leq j < m$}{
    \SetKwFunction{entropyEstimator}{\textsc{entropyEstimator}}
    $\mathbf{h}_j=$\entropyEstimator{$\mathbf{D}_{:,j}$}\\
  }
  Compute the optimal observation time distribution\\
  $j^*=\argmax_{j}\mathbf{h}_j$\\
  $\tau^*=j^* / (m - 1)$\\
  $\mathbf{t}^* = \tau^* \cdot \mathbf{p} + t_0$\\
  \Return $\mathbf{t}^*$
  \caption{Optimal observations scheduling.}
  \label{alg:optimal_scheduling}
\end{algorithm}

\subsection{Analysis of our scheduling methodology}

The advantages of our optimal scheduling methodology of binary stars observations can be summarized in the following main points:
\begin{itemize}

\item \textbf{\underline{Inference-Efficiency}:} A naive resolution of the Bayesian optimal experiment design approach involves estimating the posterior distribution incorporating virtual sampĺes at each of the candidate times $\tau_i,i\in[0,1]$ to evaluate their informational contribution to the determination of the system. This approach limits the set of candidate times to evaluate to a few due to the high computational costs involved in the Bayesian inference procedure. In contrast, our proposed maximum entropy sampling approach requires only one estimation of the original posterior distribution (without virtual samples), where its projection onto the observation space at each of the candidate times is used to assess their informational contribution.

\item \textbf{\underline{Uncertainty estimation-efficiency}:} The maximum entropy sampling criterion reduces the optimal time observation estimation problem to computing the entropy in the observation space $Y$. This computation is at most 4-dimensional (considering the 2-dimensional positional/astrometric observations and the radial velocity/spectroscopic observations of both the primary and secondary the binary system), instead of computing the entropy in the orbital parameters space $\Theta$ that has tens of dimensions (see Appendix~\ref{App:orbital_models}). Moreover, the orbital parameters space increases rapidly with the system´s multiplicity in the case of hierarchical stellar systems. Hence, performing the entropy estimation in a low-dimensional observation space reduces the computational costs dramatically, and more importantly, it simplifies the estimation problem by avoiding phenomena associated to the curse of dimensionality.

\item \textbf{\underline{Information-representativeness}:} The computational efficiency of our methodology discussed above allows us to perform the estimation of the marginal observation entropy in a dense array of normalized times $\tau\in[0,1]$, allowing to compute a pseudo-continuous curve of the marginal observation entropy over time (see top right panel on  Figures~\ref{fig:Optimal/YSC132AaAb_entropy}, \ref{fig:Optimal/HIP99675_entropy}, and \ref{fig:Optimal/LHS1070_entropy} in Section~\ref{Sec:4}). This uncertainty curve allows us not only the selection of the most informative normalized time (implementing Equation~(\ref{eq:opt_obs_time})), but it also provides a complete temporal characterization of the information gain of future observations. This valuable income can be used, e.g., to compare the information-gap between different possible observations, allowing to prioritize between them (see lower panels on Figures~\ref{fig:Optimal/YSC132AaAb_entropy}, \ref{fig:Optimal/HIP99675_entropy} and \ref{fig:Optimal/LHS1070_entropy}). Note that we can perform the same analysis even in the presence of observational constraints (e.g., due to Sun or telescopic/technique constraints) simply by removing the time segments in the entropy curve where observations can not be performed.

\item \textbf{\underline{Sequential decision}:} All the attributes mentioned in the previous points could be used to implement a sequential observational planning strategy. If we have a sequence of observation times at our disposal, we could sequentially select them. Furthermore, the characterization of the information gain of future observations can be used as a criterion to stop the process of acquiring new samples. For example, if the information gain of the best selection (after implementing the decision in Equation~(\ref{eq:opt_obs_time})) is below a certain pre-specified threshold, this could be interpreted as a condition where extra measurements do not increase the inference of the orbital parameters of the system and become thus superfluous.
\end{itemize}

Finally, it is important to mention that the optimal Bayesian design approach, described in Section~\ref{Sec:bed}, assumes that there is no direct dependency between $\theta$ and the index of the experiment yet to be selected ($\xi \in \Xi$) (this assumption is employed in Equation~(\ref{eq:bayes_mes_interchanged}) in Section~\ref{Sec:mes}). By normalizing the problem and making it tractable, we can see from the normalized time construction that this quantity depends partially on $\theta$ (since it depends on the orbital period $P$). It is worth mentioning that this observation does not limit the practical adoption of our strategy, which shows remarkable results. In theory, the gap between our normalized setting in (\ref{eq:opt_obs_tau}) and the original formulation is zero if $P$ is precisely determined. This could justify near-optimal results when $P$ is estimated with good precision, which is observed in two of the case studies explored in this work. Finally, in the regime when $P$ is not precisely estimated, the theoretical impact of this normalization strategy is a relevant topic of research that will be explored in future work.

\section{Experimental Validation}
\label{Sec:4}

The proposed scheme for estimating the optimal observation time is applied to three binary systems: the double-line visual-spectroscopic binary system HIP 89000, the single-line visual-spectroscopic binary system HIP 99675, and the visual triple hierarchical system LHS 1070. These examples are not intended to provide an exhaustive list of possible applications, but rather to give a glimpse of the performance of our methodology. The following experiments only consider the uncertainty in the orbit space ($X,Y$) since the good coverage and precision of the available RV observations turns the RV posterior uncertainty negligible.

The marginal observation entropy (in the orbit space) $H(X,Y|\mathcal{D},\tau)$ is computed along the valid range of normalized times $\tau\in[0,1]$ (with $t_0$ as the epoch of the latest available observation), using the k-nearest neighbour entropy estimation method \citep{kozachenko1987sample} on $1000$ randomly selected samples of the posterior distribution $\theta\sim p(\Theta|\mathcal{D})$. These samples were estimated through the Markov chain Monte Carlo algorithm No-U-Turn sampler \citep{hoffman2014no}, previously presented in \citet{videla2022bayesian}. Following the experimental setting of that work, we simulate 4 independent Markov chain from starting points determined by the quasi-Newton optimization method L-BFGS \citep{liu1989limited}.

We perform a comparative analysis between the marginal observation entropy $H(X,Y|\mathcal{D},\tau_i)$ and the posterior predictive distributions $p(X,Y|\mathcal{D},\tau_i)$ at five representative normalized times $\tau_i,i\in\{1,2,3,4,5\}.$
%\footnote{While one may naively consider that in order to determine the seven basic orbital elements of a visual binary we need at least seven observational points, in reality, the orbital coverage is more critical. Also, in our case, the analyzed systems already have some historical data, and the task is to evaluate when new observations should be placed in time in order to significantly improve the orbit. Note also that in our analysis we disregard any other constrains imposed by the actual observability of the targets.}.
Additionally, we estimate the updated posterior distribution $p(\Theta|\mathcal{D}\cup\{d_{\tau_i}\})$ by incorporating virtual positional observations $d_{\tau_i}$ (and redoing the inference process through the No-U-Turn sampler algorithm) at the same representative normalized times. Each virtual observation corresponds to the position of the maximum a posteriori orbit (from the maximum a posteriori orbital parameter $\theta_{map}$) at the epochs $t_i^{map}=\tau_i\cdot  P_{map}+t_0$, i.e., $d_{\tau_i}=(f_{\text{kep}}^{x}(\theta_{map},t_i^{map}),f_{\text{kep}}^{y}(\theta_{map},t_i^{map}))$, with $\sigma_i=\sigma_{min}$, where $\sigma_{min}$ is the minimum standard deviation from the available observations $\mathcal{D}$. To validate our methodology, we compare the marginal observation entropies $H(X,Y|\mathcal{D},\tau_i)$ with the respective updated posterior distributions $p(\Theta|\mathcal{D}\cup\{d_{\tau_i}\})$.

Finally, we present the estimated distribution of the optimal epoch of observation $t^*$ associated to the normalized time of highest marginal observation entropy $\tau^*=\argmax_{\tau\in [0,1]} H(Y|\mathcal{D},\tau)$.

\subsection{HIP 89000}
\label{Sec:5.1}

\begin{figure}[!h]
    \centering
    \includegraphics[width=\textwidth]{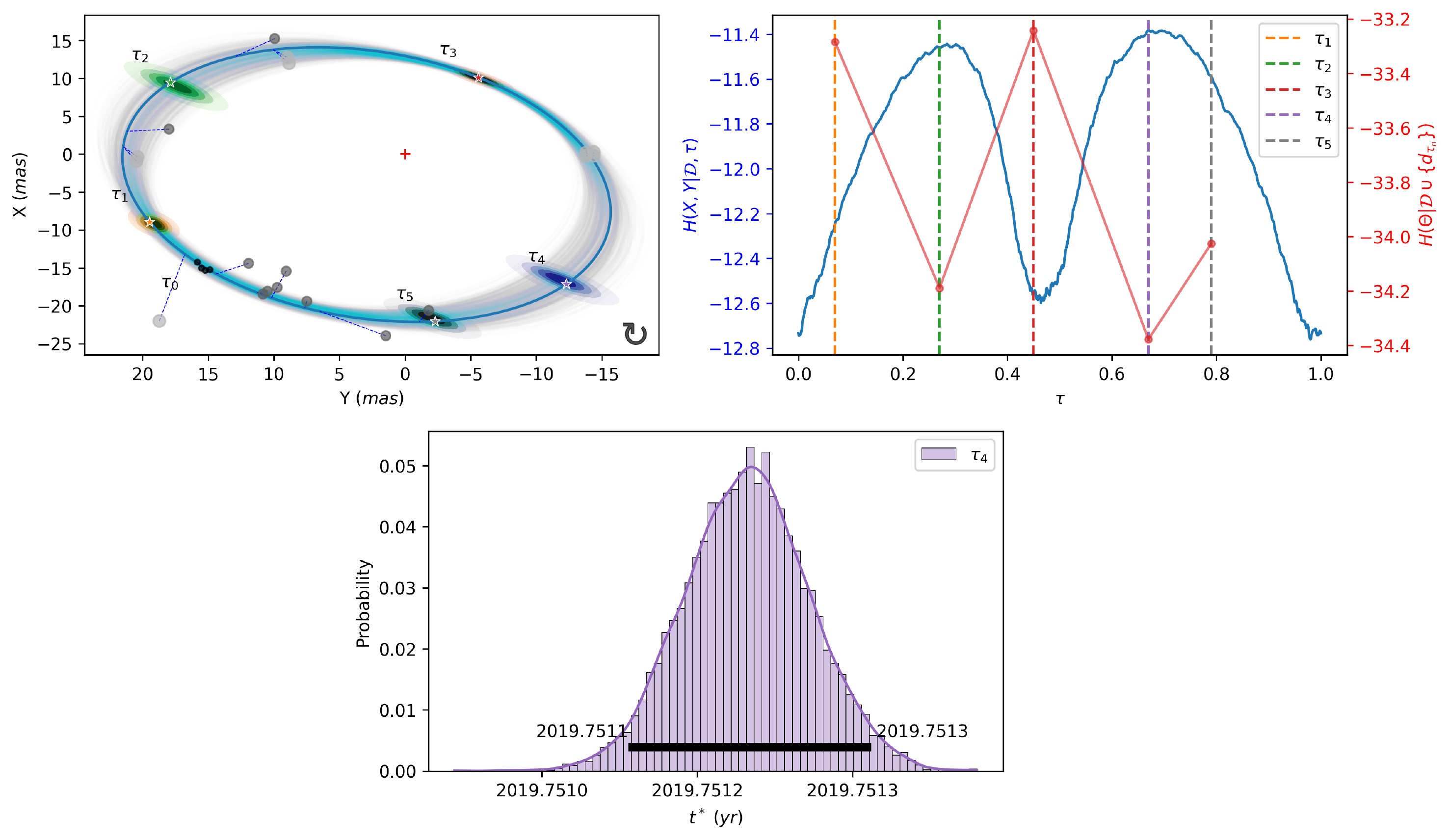}
\caption{Optimal observation time analysis on the binary system HIP 89000. Top left: Posterior predictive distribution $p(X,Y|\mathcal{D})$ along $\tau\in[0,1]$ (represented by 1000 cyan orbits), highlighting the projection at the selected representative epochs $\tau_i,i\in\{1,2,3,4,5\}$ (color clouds). The star of each color cloud represents the projection of the selected epochs $\tau_i$ in the MAP orbit (blue line). The actual interferometric observations are represented by grey dots, with a size and darkness proportional to their declared uncertainties (smaller and darker for smaller uncertainties). These are joined by dashed lines to their predicted position on the MAP orbit. 
Top right: Estimated marginal observation entropy curve $H(X,Y|\mathcal{D},\tau)$ (left ordinate) and estimated updated posterior distribution entropy $H(\Theta|\mathcal{D}\cup \{d_{\tau_i}\})$ (right ordinate) when incorporating the virtual observations $d_{\tau_i}$ at the five representative normalized times $\tau_i$, which are highlighted by the vertical dashed lines. Bottom: Distribution of optimal observation time induced by the normalized time of highest entropy in the observation space $\tau^*=\argmax_{\tau\in[0,1]}H(X,Y|\mathcal{D},\tau)$, in this case $\tau_4$. The vertical black line at the bottom of the histogram indicates the 95\% high posterior density interval HPDI (the
narrowest interval that contains 95\% of the posterior distribution, including the mode).}
\label{fig:Optimal/YSC132AaAb_entropy}
\end{figure}

The system HIP 89000 (discovery designation YSC132AaAb) is an $SB2$ binary presented and solved most recently by \citet{videla2022bayesian}. The available data consists of interferometric observations mostly concentrated around the apoastron passage (the binary is very tight), but with abundant and precise observations of RV of both components. The astrometric observations and their errors are visualized in Figure~\ref{fig:Optimal/YSC132AaAb_entropy} (top left panel). The estimated marginal observation entropy curve $H(X,Y|\mathcal{D},\tau),\tau\in[0,1]$, the posterior predictive distribution $p(X,Y|\mathcal{D})$, their respective projections at the five selected normalized times $\tau_i,i\in\{1,2,3,4,5\}$, and the estimated distribution of the optimal observation time $t^*$ are presented in Figure~\ref{fig:Optimal/YSC132AaAb_entropy} as well.

The estimated entropy curve shows two local maxima with a prominent local minimum value in between, and two minima at the beginning $\tau=0$ and at the end $\tau=1$ of the curve. The behavior of the estimated entropy curve coincides with the observed dispersion of the posterior distribution projected in the observation space, where the times of minimum entropy are the corresponding zones in the orbit populated with high precision observations ($\tau=0$), presenting a narrow band of the projected orbits. The entropy increase with $\tau$ as well as the dispersion of the projected orbits as no precise observations are presented in that zone of the orbital space ($\tau_1$). This behavior persists until reaching the first local maximum of the entropy, where the dispersion of the projected orbits is highest ($\tau_2$) due to the scarce and low precision observations in that zone of the orbit. Then, the entropy starts to decrease as well as the dispersion of the projected orbits until reaching a local minimum value ($\tau_3$). It is interesting to note that the projected orbits becomes narrower (with lower uncertainty) even in the absence of positional observations near that zone, attributed to an underlying uncertainty reduction of some of the orbital parameters through observations in other zones of the orbit. This result show that the naive idea that observations should be always placed were there are no previous observations is not necessarily the most efficient thing to do. The estimated entropy curve increase again until reaching a second local maximum ($\tau_4$), which is slightly higher than the previous local maximum (at $\tau_2$). The projected orbits reach the highest dispersion due to the total absence of observations in that zone of the orbit. Finally, the entropy curve starts to decrease at the end of the observational period ($\tau=1$) reaching a similar minimum as the beginning of the curve ($\tau=0$) due to the periodicity of the orbit.

\begin{figure}[!h]
    \centering
    \includegraphics[width=\textwidth]{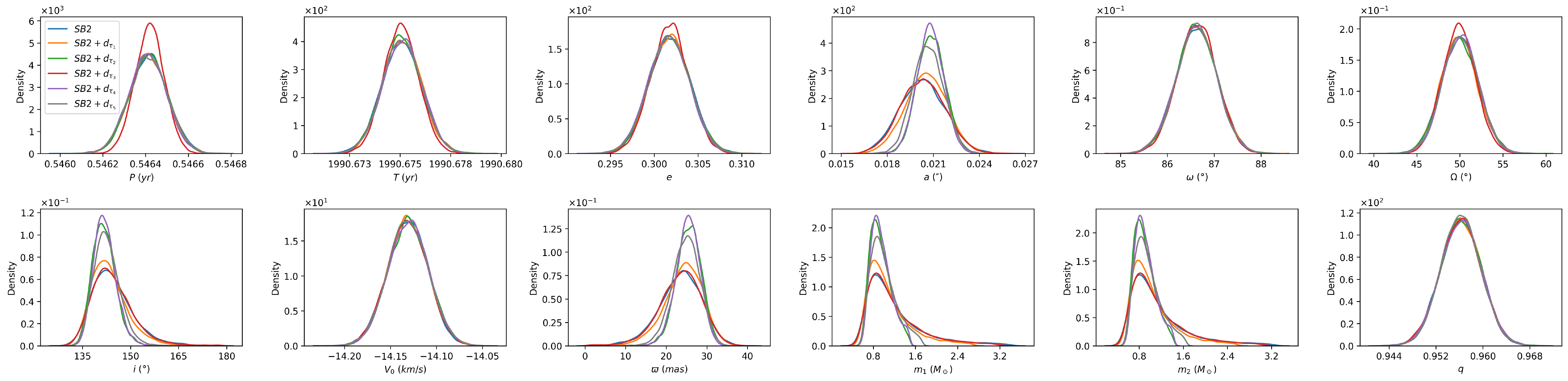}
    \caption{Effect on the marginal posterior distributions for the orbital elements and derived physical parameters of the binary HIP 89000 after incorporating virtual observations $d_{\tau_i}
    \sim\mathcal{N}(f_{\text{kep}}(\theta_{map},\tau_i),\sigma^2),i\in\{1,2,3,4,5\}$, with $\sigma^2$ fixed to the minimum variance of the available system's observations $\mathcal{D}$. The impact and relevance of acquiring data at different epochs can be clearly appreciated in these plots, being in consistency with the predictions from the top right panel on Figure \ref{fig:Optimal/YSC132AaAb_entropy}, i.e., the impact in reducing parameter uncertainty is largest when the observations are acquired in regions of large marginal observation entropy.}
    \label{fig:Optimal/YSC132AaAb_params}
\end{figure}

\begin{figure}[!h]
    \centering
    \includegraphics[width=\textwidth]{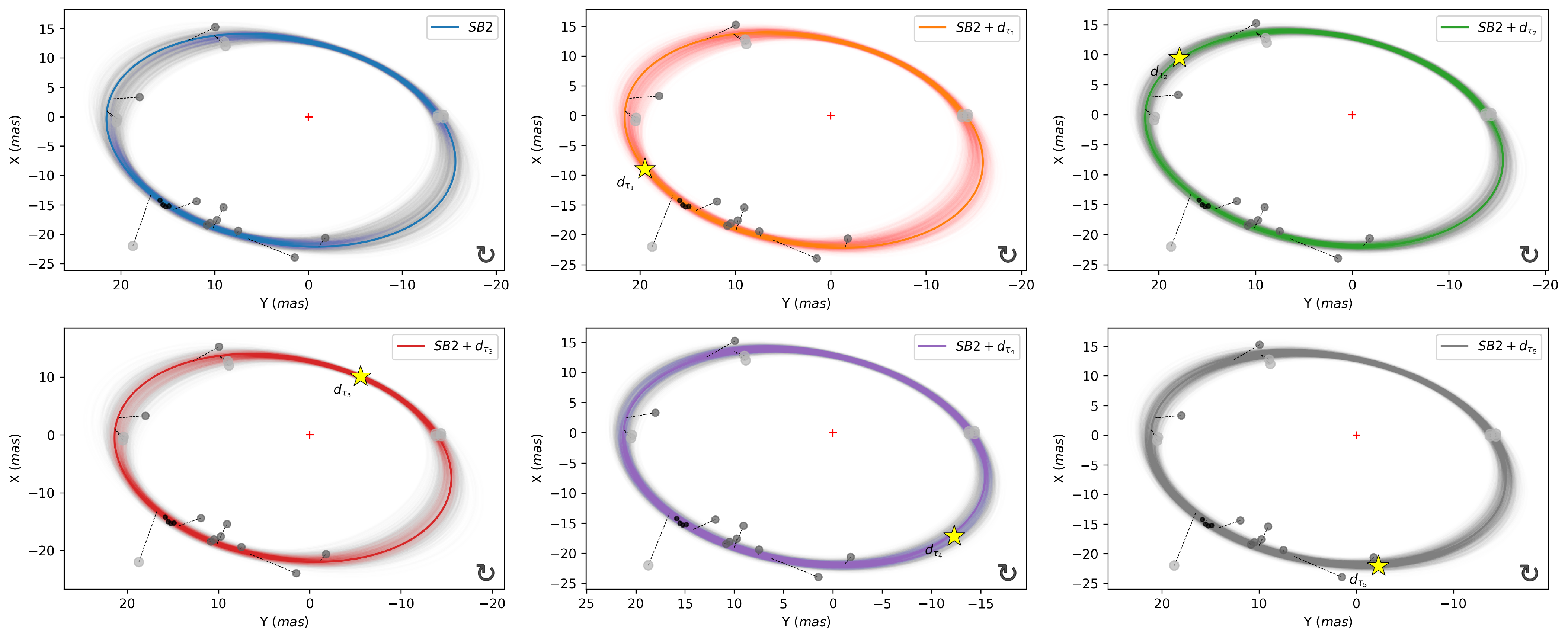}
    \caption{Projection on the orbital configuration 
    %(upper row) and radial velocity curve (lower row)
    from the updated posterior predictive distribution for the system HIP 89000 when incorporating the virtual observations $d_{\tau_i}$ at epochs $\tau_i,i\in\{1,2,3,4,5\}$ (indicated by the yellow star). For numerical values of the orbital parameters, please see Table \ref{tab:YSC132AaAb_params} in Appendix \ref{App:complementary_results}.}
    \label{fig:Optimal/YSC132AaAb_observtions}
\end{figure}

The marginal updated posterior distributions after the incorporation of each sample $d_{\tau_i}$ are presented in Figure \ref{fig:Optimal/YSC132AaAb_params}, while corresponding projections on the observational space are presented in Figure \ref{fig:Optimal/YSC132AaAb_observtions}. The comparison of the marginal updated posterior distributions show different magnitudes on the uncertainty of the orbital parameters (dispersion of the distributions), where the virtual observations $d_{\tau_i}$ that yield the highest reduction in uncertainty of orbital parameters space are those in which the corresponding entropy $H(X,Y|\mathcal{D},\tau_i)$ is higher and vice-versa. This behavior is clearly reflected in the comparison between the estimated values of $H(X,Y|\mathcal{D},\tau_i)$ and $H(\Theta|\mathcal{D}\cup\{d_{\tau_i}\})$ presented in the top right panel of the Figure \ref{fig:Optimal/YSC132AaAb_entropy}, confirming the equivalence between the maximum entropy sampling criterion and the Bayesian optimal design problem. The projected updated posterior distributions show a considerable uncertainty reduction when the virtual measurement is taken in the highest uncertainty zones. On the other hand, the uncertainty reduction is negligible when the virtual measurements are acquired in zones of lower uncertainty, satisfying the objective of the maximum entropy sampling principle. Notably, the updated marginal posterior distribution $SB2+d_{\tau_3}$ (red distribution in Figure \ref{fig:Optimal/YSC132AaAb_params}) presents the lowest dispersion on the orbital parameters $P$, $T$, $e$, and $\Omega$, since the associated virtual observation $d_{\tau_{3}}$ (red star on top left panel of Figure \ref{fig:Optimal/YSC132AaAb_entropy}) absorbs the uncertainty exhibited by the corresponding posterior predictive distribution in $\tau_3$ (red cloud on top left panel of Figure \ref{fig:Optimal/YSC132AaAb_entropy}). Unlike all the other cases, this projected uncertainty is dispersed along the orbit, mainly due to the orbital period uncertainty, and hence, simulating a virtual observation on that area of the orbit causes the greatest reduction on the mentioned parameters. However, the updated marginal posterior distribution $SB2+d_{\tau_3}$ also exhibits no uncertainty reduction in the other orbital parameters compared to the original posterior distribution $SB2$ (red and blue distributions in Figure \ref{fig:Optimal/YSC132AaAb_params}, respectively), and hence, the main orbital zones of uncertainty ($\tau_2$ and $\tau_4$) remain unchanged (red orbit in Figure \ref{fig:Optimal/YSC132AaAb_observtions}). We conclude that observing at $\tau_3$ provides the lowest information gain for the studied system, as seen in the red-doted line of the entropy curve in the top right panel of Figure \ref{fig:Optimal/YSC132AaAb_entropy}.

Finally, the distribution of the optimal observation time $t^*$ at the normalized time of maximum uncertainty of the observation space ($\tau_4$) is presented in the lower panel of Figure \ref{fig:Optimal/YSC132AaAb_entropy}. The estimated distribution has a clearly defined Gaussian shape with extremely low dispersion. This result reflects that the period $P$ of the stellar system is well determined due to the presence of abundant and precise radial velocities observations.

\subsection{HIP 99675}
\label{Sec:5.2}

\begin{figure}[!h]
    \centering
    \includegraphics[width=\textwidth]{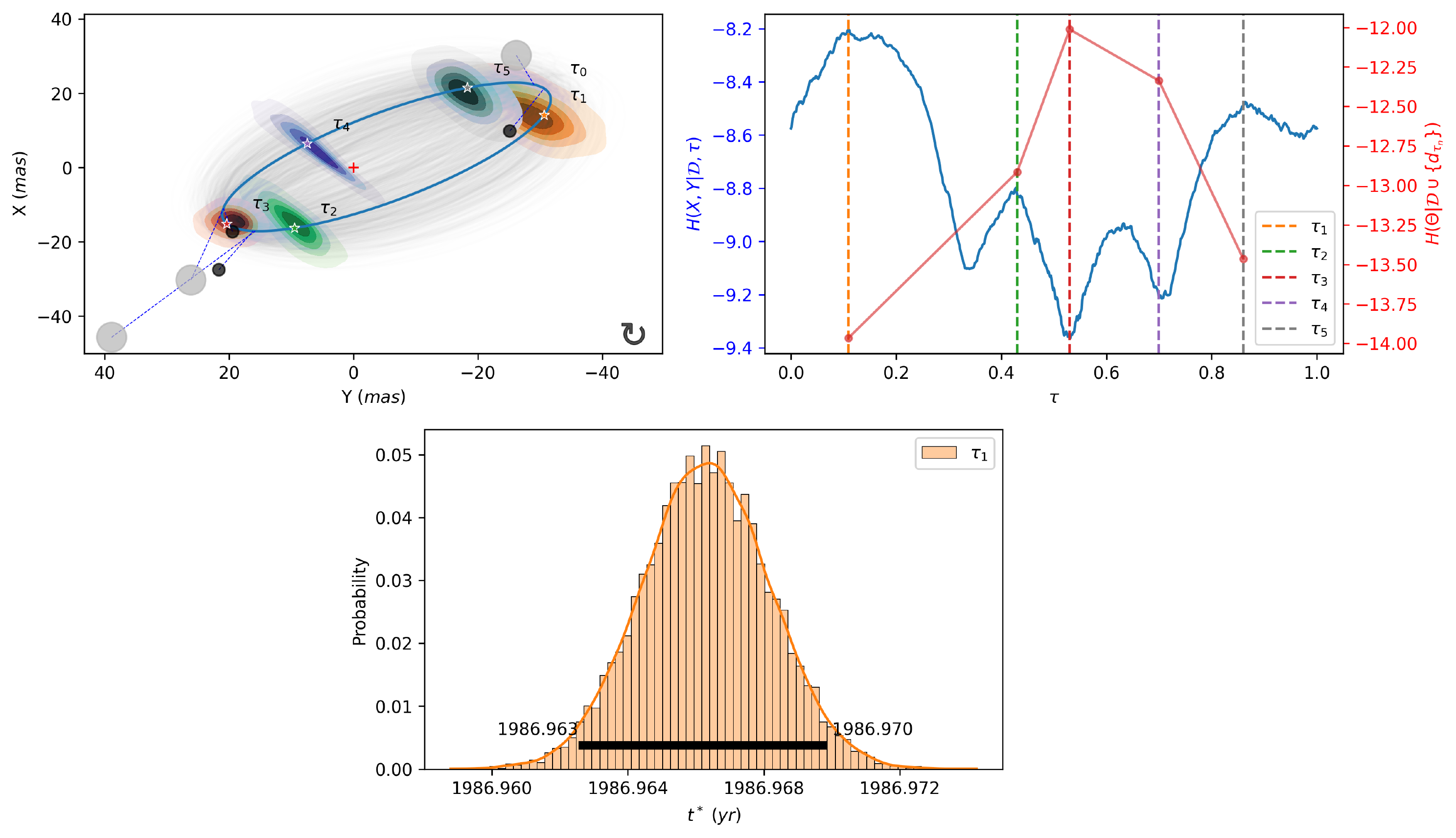}
\caption{Same as Figure~\ref{fig:Optimal/YSC132AaAb_entropy} but for the stellar system HIP 99675.}
    \label{fig:Optimal/HIP99675_entropy}
\end{figure}

The system HIP 99675 (discovery designation WRH33Aa,Ab) is a $SB1$ binary presented and solved most recently by \citet{videla2022bayesian}. The data for this object consists of few astrometric observations in two extreme zones of the orbit, and abundant and highly precise RV observations of the primary component. The observations and their errors are visualized in Figure~\ref{fig:Optimal/HIP99675_entropy} (left panel, this is also a tight binary that require interferometric measurements). The estimated entropy marginal observation entropy curve $H(X,Y|\mathcal{D},\tau),\tau\in[0,1]$, the posterior predictive distribution $p(X,Y|\mathcal{D})$, their respective projections at the five selected normalized times $\tau_i,i\in\{1,2,3,4,5\}$, and the estimated distribution of the optimal epoch of observation $t^*$ are also presented in Figure~\ref{fig:Optimal/HIP99675_entropy}.

The estimated entropy curve has a quasi-symmetrical behavior around the point of minimum entropy ($\tau_3$), where the entropy increases away from that point ($\tau<\tau_3$ and $\tau>\tau_3$) in a wavy manner, presenting multiple local minima and maxima but permanently preserving its average increasing behavior. The behavior of the estimated entropy curve apparently does not relate to the observed posterior orbits that form a diffuse cloud of lines, with no clearly defined boundaries due to the scarce positional observations. However, the projection of the posterior distribution at the five normalized times selected from the estimated entropy curve $\tau_i,i\in\{1,2,3,4,5\}$ shows that the uncertainties are mostly perpendicularly distributed with respect to the well-defined orbits, and the overlap of these orbits give origin to the diffuse cloud. The zone in the orbit space with the lowest entropy ($\tau_3$) coincides with most of the available positional observations being near that zone. In contrast, the opposite zone in the orbit presents the highest entropy ($\tau_1,\tau_5$). The wavy behavior of the entropy curve is reflected in the shape of the projection of the posterior distribution in the orbit space, where local minima correspond to long but narrow projections ($\tau_4$) while local maxima corresponds to slightly shorter but wider projections ($\tau_2$).

\begin{figure}[!h]
    \centering
    \includegraphics[width=\textwidth]{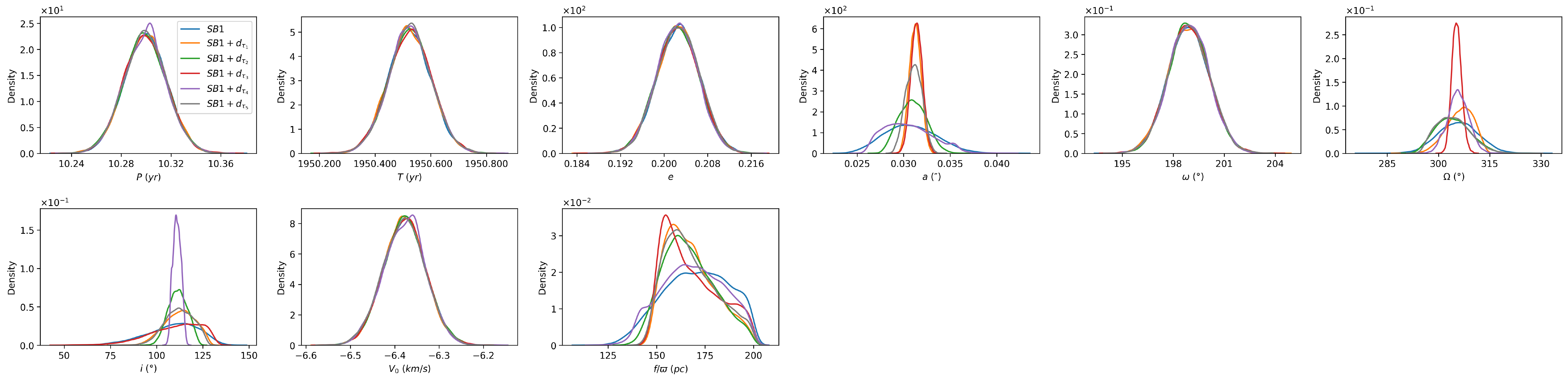}
    \caption{Same as Figure~\ref{fig:Optimal/YSC132AaAb_params} but for the stellar system HIP 99675. Since this is a single line spectroscopic binary, we can not compute orbital parallaxes nor the individual stellar masses as done for HIP~89000, instead we can compute the ratio $f/\varpi$ as defined by \citet{videla2022bayesian} (see Section~2.2 and 2.4 on that paper).}
    \label{fig:Optimal/HIP99675_params}
\end{figure}

\begin{figure}[!h]
    \centering
    \includegraphics[width=\textwidth]{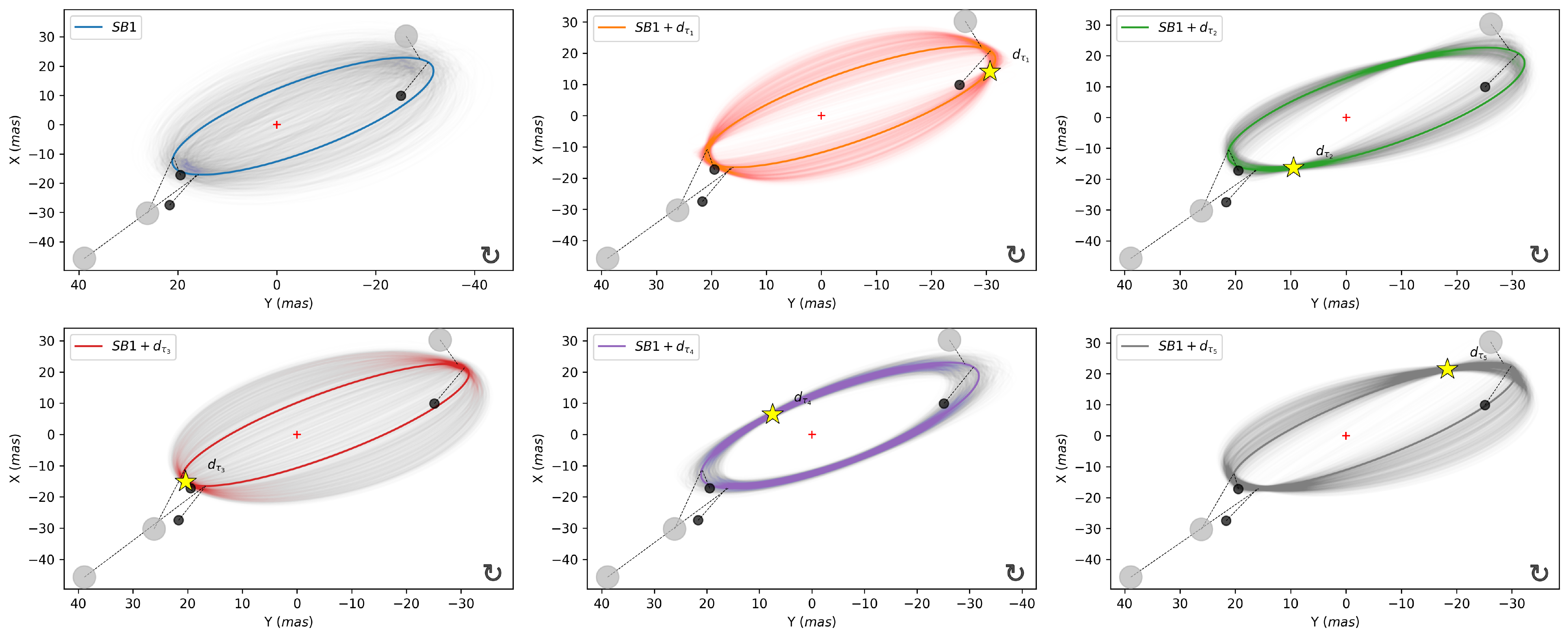}
    \caption{Same as Figure~\ref{fig:Optimal/YSC132AaAb_observtions} but for the stellar system HIP 99675.  For numerical values of the orbital parameters, please see Table \ref{tab:HIP99675_params} in Appendix \ref{App:complementary_results}.}
    \label{fig:Optimal/HIP99675_observtions}
\end{figure}

The marginal updated posterior distributions obtained by incorporating each sample $d_{\tau_i}$ are presented in Figure \ref{fig:Optimal/HIP99675_params} and the corresponding projections on the observation space are presented in Figure \ref{fig:Optimal/HIP99675_observtions}. The posterior distribution updated with the virtual observation generated through the normalized time of lowest entropy ($d_{\tau_3}$) presents an uncertainty reduction on some of the marginal distributions of the parameters ($a,\Omega$), but a considerable spread  (i.e., a lack of uncertainty reduction) in other orbital parameters ($i$), compared to the other updated posterior distributions. This trade-off between the uncertainty reduction/increase between some of the orbital parameters is observed in each case, inducing different updated posterior predictive distributions. We note that the incorporated virtual observations reduce the projected uncertainty in the zone of the orbit space near that measurement, as expected. For example, the $d_{\tau_3}$ observation highly reduce the uncertainty on the length of the orbit, coinciding with a reduction on the uncertainty of the updated marginal posterior distributions for the semi-major axis $a$ and $\Omega$. However, this virtual observation does not provide any relevant information about the inclination of the orbit, expressed by a large scatter on the updated marginal posterior distributions of $i$ (by contrast, as it can be seen on Figure~\ref{fig:Optimal/HIP99675_params}, an observation at $d_{\tau_4}$ will help pin-point $i$ very accurately). The computed entropy of the updated posterior distribution $H(\Theta|\mathcal{D}\cup\{d_{\tau_i}\})$ takes into account the aforementioned trade-off between the increase and reduction of uncertainty of the posterior distribution along its different dimensions.

Once again, as it was the case for HIP 89000, the comparison between the entropies $H(X,Y|\mathcal{D},\tau_i)$ and $H(\Theta|\mathcal{D}\cup\{d_{\tau_i}\})$ presented in the top right panel of Figure~\ref{fig:Optimal/HIP99675_entropy} shows that the equivalence between the maximum entropy sampling criterion and the Bayesian optimal design problem is preserved.

The distribution of the optimal observation time $t^*$ at the normalized time of maximum uncertainty of the observation space ($\tau_1$) is presented in the lower panel of Figure \ref{fig:Optimal/HIP99675_entropy}. Similarly to the case of HIP 89000, the estimated distribution has a clearly defined Gaussian shape with an extremely low dispersion, which reflects that the period $P$ of the stellar system is well determined due to the presence of abundant and precise radial velocities observations, as it was also the case for HIP 89000.

\subsection{LHS 1070}
\label{Sec:5.3}

\begin{figure}[!h]
    \centering
    \includegraphics[width=\textwidth]{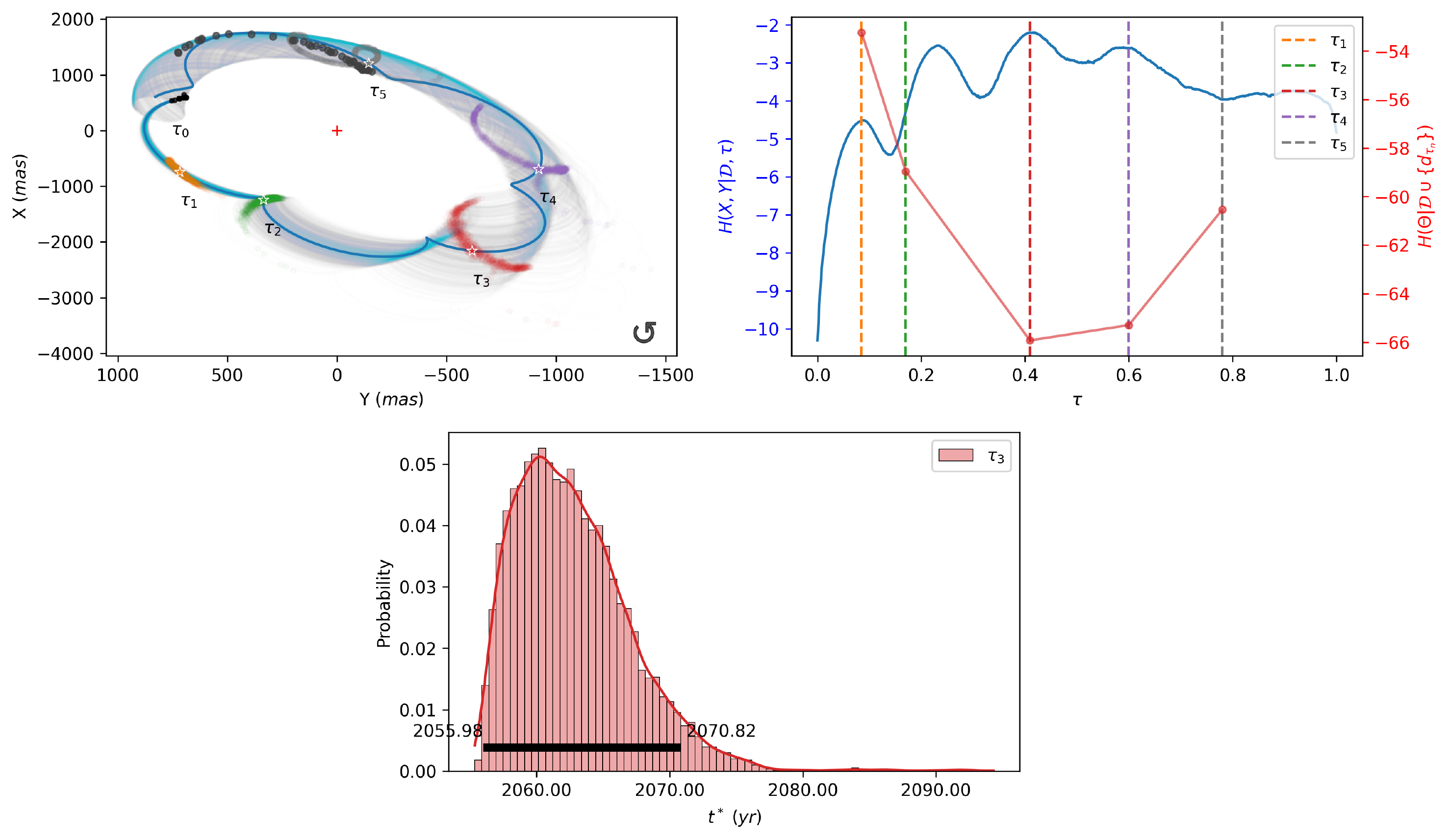}
\caption{Same as Figure~\ref{fig:Optimal/YSC132AaAb_entropy} but for the stellar system LHS 1070.}
    \label{fig:Optimal/LHS1070_entropy}
\end{figure}

The system LHS 1070 is an hierarchical triple stellar system presented and solved most recently by \citet{villegas2021bayes}. This hierarchical system is composed of tight outer pair BaBb that orbits a brighter primary component A. The available data consists of numerous and precise positional observations of the pair BaBb that cover the entire orbit, and precise observations of the brighter component of the outer pair (Ba) relative to the primary component A. Given its relative faintness ($V=15.3$) no radial velocity observations are available for any of the system´s components. The observations and their errors are visualized in Figure~\ref{fig:Optimal/LHS1070_entropy} (top left). The estimated entropy marginal observation entropy curve $H(X,Y|\mathcal{D},\tau),\tau\in[0,1]$, the posterior predictive distribution $p(X,Y|\mathcal{D})$, their respective projections at the five selected normalized times $\tau_i,i\in\{1,2,3,4,5\}$, and the estimated distribution of the optimal observation time $t^*$ are presented in Figure~\ref{fig:Optimal/LHS1070_entropy}.

The estimated entropy curve presents a global minimum at the beginning $\tau=0$, which coincides with the zone in the orbit populated with abundant and precise positional observations. The entropy curve increases steeply from the positional observations in the orbit space until reaching a maximum value at $\tau_3$. The rest of the curve presents a constant value with a slight decrease starting after $\tau_4$. The entropy curve shows an oscillating behavior along all the normalized times $\tau$ evaluated. This behavior coincides with the lobes of the projected orbits in the observation space. The projection of the posterior distribution in the orbit space at the normalized times $\tau_i,i\in\{1,2,3,4,5\}$ shows that the posterior distributions increase its dispersion in zones far from the available observations, forming wavy patterns. This pattern can be attributed to the shift increase between the projected orbits due to the large uncertainty on the orbital period of the purely astrometric hierarchical system (as mentioned, no radial velocities are available for this system).

\begin{figure}[!h]
    \centering
    \includegraphics[width=\textwidth]{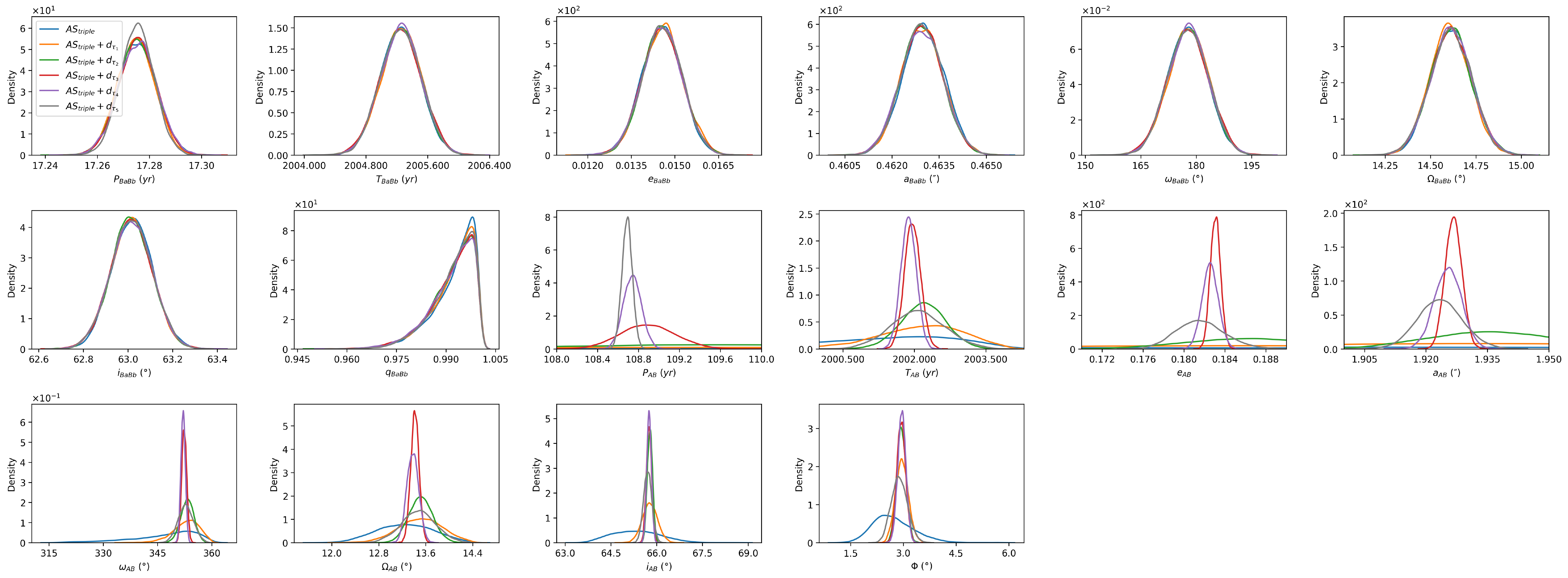}
    \caption{Same as Figure~\ref{fig:Optimal/YSC132AaAb_params} but for the stellar system LHS 1070. Being this a triple stellar system, in addition to the classical orbital elements, the rightmost figure in the bottom panel shows the mutual inclination between the two orbits (see Equation~(\ref{eq:mutual})), which is often considered in the study of these systems.}
    \label{fig:Optimal/LHS1070_params}
\end{figure}

\begin{figure}[!h]
    \centering
    \includegraphics[width=\textwidth]{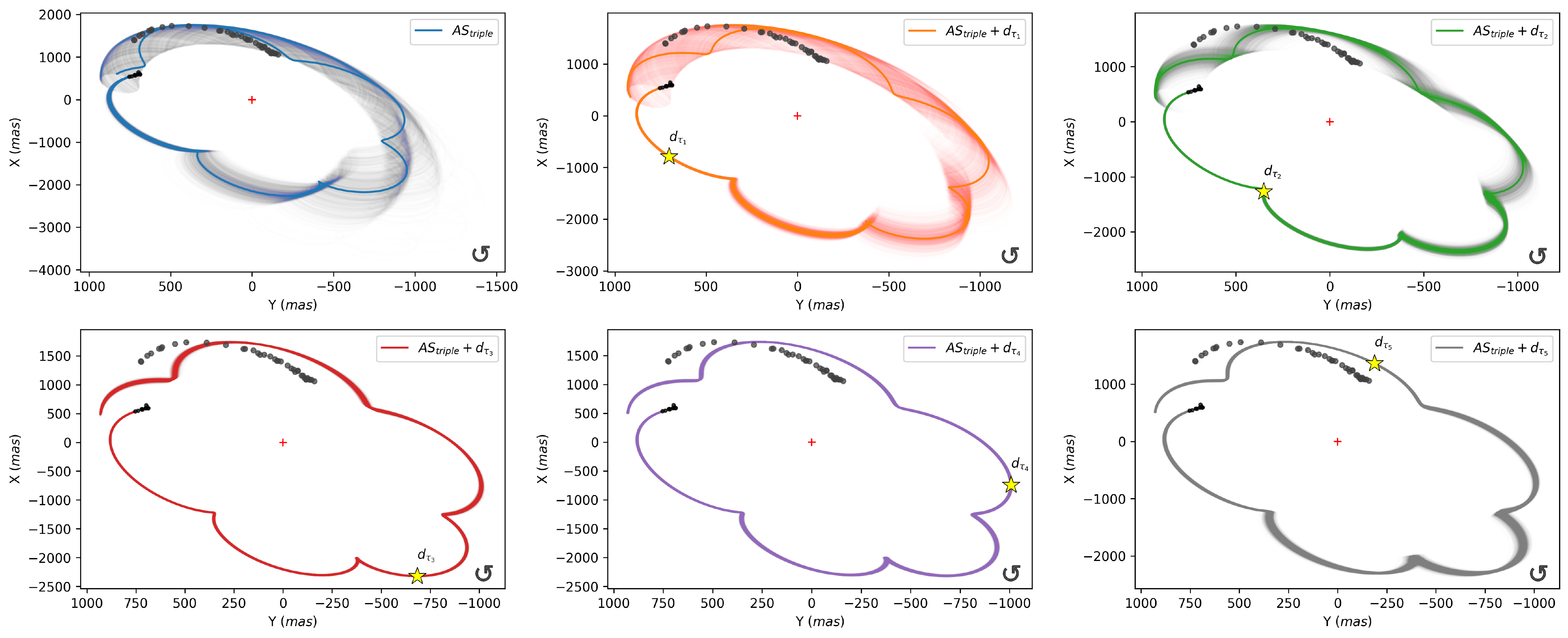}
    \caption{Same as Figure~\ref{fig:Optimal/YSC132AaAb_observtions} but for the stellar system LHS 1070.  For numerical values of the orbital parameters, please see Table \ref{tab:LHS1070_params} in Appendix \ref{App:complementary_results}. The divergence of the orbit with respect to the measured position is due to the fact that these orbits are not closed and, as indicated in the second paragraph at the beginning of  Section~\ref{Sec:4}, the orbit is computed from the last observation in the range $\tau\in[0,1]$. Note that this is also seen in Figure~\ref{fig:Optimal/LHS1070_entropy} (top left panel).}
    \label{fig:Optimal/LHS1070_observtions}
\end{figure}

The marginal updated posterior distributions obtained by incorporating each sample $d_{\tau_i}$ are presented in Figure~\ref{fig:Optimal/LHS1070_params}, while the corresponding projections on the observation space are presented in Figure~\ref{fig:Optimal/LHS1070_observtions}. The comparison of the marginal updated posterior distributions shows different magnitudes of uncertainty of the orbital parameters of the outer orbit $AB$, where the virtual observations $d_{\tau_i}$ that reaches the highest uncertainty reduction in the orbital parameters space are the observations in which the corresponding entropy $H(X,Y|\mathcal{D},\tau_i)$ is the highest and viceversa. This opposite behavior is clearly reflected in the comparison between the estimated values of $H(X,Y|\mathcal{D},\tau_i)$ and $H(\Theta|\mathcal{D}\cup\{d_{\tau_i}\})$ presented in the top right panel of the Figure \ref{fig:Optimal/LHS1070_entropy}, which again verifies the equivalence between the maximum entropy sampling criterion and the Bayesian optimal design problem. The updated posterior distributions show a considerable uncertainty reduction in the sections of the orbit where the virtual measurement is obtained in the zones of highest uncertainty. On the other hand, the uncertainty reduction is negligible when the virtual measurement are taken in those zones of lower uncertainty, once again satisfying the objective of the maximum entropy sampling principle. We  note that the updated posterior predictive distributions (in Figure~\ref{fig:Optimal/LHS1070_observtions}) shows that incorporating a virtual observation on the first orbital lobe ($AS_{triple}+d_{\tau_{1}}$ case) not only absorbs the uncertainty of that lobe, but also most of the uncertainty of the next lobe (in the sense of rotation). Similarly, incorporating a virtual observation in the intersection of the first two orbital lobes ($AS_{triple}+d_{\tau_{2}}$ case) decreases significantly the uncertainty for both of them, and a considerable amount of the uncertainty of the third lobe. Remarkably, incorporating a virtual observation in the third orbital lobe ($AS_{triple}+d_{\tau_{3}}$ case) absorbs almost all the uncertainty of all the five orbital lobes, being the case with biggest uncertainty reduction of the whole system, as predicted by the entropy curve (top right panel of Figure~\ref{fig:Optimal/LHS1070_entropy}). Hereinafter, we observe that incorporating virtual observations in the following orbital lobes ($SB2+d_{\tau_{4}}$ and $AS_{triple}+d_{\tau_{5}}$ cases) effectively absorbs the uncertainty of the last portion of the orbit, but does not allow to absorb the uncertainty of the first orbital lobes, as in the previous cases. We conclude that virtual observations at $d_{\tau_{4}}$ and $d_{\tau_{5}}$ are less informative than $d_{\tau_{3}}$, as expected.

The distribution of the optimal observation time $t^*$ at the normalized time of maximum uncertainty of the observation space ($\tau_3$) is presented in the lower panel of the Figure \ref{fig:Optimal/LHS1070_entropy}. Unlike the previously studied cases, the estimated distribution presents a high dispersion, which reflects the high uncertainty on the period $P$ of the outer system $AB$ due to the low coverage of positional observation in orbit space, and the absence of radial velocity measurements. In this high uncertainty regime for $P$, our normalized scheme can still select a very informative sample: a sample that significantly reduces the posterior entropy of the parameters. This shows that our methodology is also suitable when the period $P$ is poorly characterized, which translates into uncertainty in the selection of $t^*$. Then, the uncertainty in  $t^*$ does not mean that selecting the most plausible candidate from this distribution is uninformative for the main estimation task.  

\section{Conclusions and outlook}
\label{Sec:conclu}

In this paper we present an application for the estimated posterior distribution of the orbital elements of stellar binary and multiple systems (first introduced in \citet{videla2022bayesian}) to the determination of the optimal observation time, i.e., the time that mostly reduces the uncertainty of the orbital parameters that characterize the system. Indeed, \citet{videla2022bayesian} have emphasized the importance of providing a complete characterization of the posterior distributions of orbital parameters on binary systems through a Bayesian perspective. This approach allows not only to procure an uncertainty quantification (as many classical optimization-based methods roughly do), but it also allows to visualize the uncertainty of the system in the observational space. This last aspect is a relevant input for other statistical tasks, such as the optimal observation problem addressed in this work.

Our Bayesian methodology based on the maximum entropy sampling principle exploits the periodic nature of binary systems, allowing us to supply a full temporal characterization of its uncertainty. Subsequently, the determination of the instant of observation that leads to the highest and lowest information about the system through the incorporation of a new observation can be addressed, providing a probability distribution of the optimal observation time.

The theoretical foundations and practical advantages of the proposed methodology have been extensively discussed. We show that the numerical efficiency of our method allows us to compute an information gain curve in a pseudo-continuous range of time that effectively characterizes the uncertainty variability of the system. The proposed framework was applied to three different stellar systems, showing the suitability and capabilities of the method for the determination of the optimal observation time.

We also show that the generated information gain curve for the stellar system allows us to conduct a deeper analysis of the uncertainty of the orbital and radial velocity curves of the systems. These curves offer valuable and interpretable probabilistic descriptions of the stellar system.

As discussed in Section~\ref{Sec:3.4}, our definition of the normalized time for the system's periodicity does pose a theoretical inconsistency when the orbital period of the studied object is not well determined. However, the experiments on the high periodic-uncertain object LHS~1070 show that our proposed methodology seems to still be effective, even in the mentioned scenario. The theoretical impact of this normalization is a relevant topic of research that will be explored in future work.

\newpage

%\begin{acknowledgments}

MV and RAM acknowledge support from FONDECYT/ANID grant No. 1190038. JS and MO acknowledge support from FONDECYT/ANID grant No. 1210315  and from the Advanced Center for Electrical and Electronic Engineering under Basal Project FB0008. RAM acknowledges the European Southern Observatory in Chile for its hospitality during a sabbatical leave in which this work was finished.

We are grateful to Andrei Tokovinin (CTIO/NOIRLab) for his help and support on the use of HRCam at SOAR, to Ricardo Salinas (GS/ZORRO instrument scientist), Elise Furlan (Scientist, NASA Exoplanet Science Institute Caltech/IPAC), and Steve Howell (scientist Space Science and Astrobiology Division, NASA Ames Research Center) for their help and support on the use of ZORRO at Gemini South. We also acknowledge all the support personnel at CTIO and GS for their commitment to operations during these difficult COVID times.

This research has made use of the Washington Double Star Catalog maintained at the U.S. Naval Observatory and of the SIMBAD database, operated at CDS, Strasbourg. We are very grateful for the continuous support of the Chilean National Time Allocation Committee under programs CN2018A-1, CN2019A-2, CN2019B-13, CN2020A-19, CN2020B-10 and CN2021B-17 for SOAR, and the Gemini Time Allocation Committee under program ID GS-2019A-Q-110, GS-2019A-Q-311, GS-2019B-Q-116, GS-2019B-Q-223, GS-2020A-Q-116, and GS-2020B-Q-142.

Some of the Observations in this paper made use of the High-Resolution Imaging instrument ZORRO. ZORRO was funded by the NASA Exoplanet Exploration Program and built at the NASA Ames Research Center by Steve B. Howell, Nic Scott, Elliott P. Horch, and Emmett Quigley. ZORRO is mounted on the Gemini South telescope of the international Gemini Observatory, a program of NSF’s NOIRLab, which is managed by the Association of Universities for Research in Astronomy (AURA) under a cooperative agreement with the National Science Foundation on behalf of the Gemini Observatory partnership: the National Science Foundation (United States), National Research Council (Canada), Agencia Nacional de Investigaci\'on y Desarrollo (Chile), Ministerio de Ciencia, Tecnolog\'{i}a e Innovaci\'on (Argentina), Minist\'erio da Ci\^{e}ncia, Tecnologia, Inova\c{c}\~{a}es e Comunica\c{c}\~{o}es (Brazil), and Korea Astronomy and Space Science Institute (Republic of Korea).

%\end{acknowledgments}

\vspace{5mm}
\facilities{{\bf CTIO:SOAR, Gemini}}

\software{Stan \citep{carpenter2017stan},  
          NumPy \citep{harris2020array}, 
          Matplotlib \citep{hunter2007matplotlib},
          Seaborn \citep{Waskom2021seaborn},
          ArviZ \citep{arviz_2019}.
          }

\newpage

\appendix

\section{Orbital Keplerian models}
\label{App:orbital_models}

\subsection{Visual-spectroscopic binary stellar system}
\label{App:binaries}

Neglecting the effects of mass transfer and complex relativistic phenomena as well as the interference of other celestial bodies, the orbit of binary stellar systems is characterized by seven orbital parameters: the \textit{time of periastron passage} $T$, the period $P$, the orbital \textit{eccentricity} $e$, the orbital \textit{semi-major axis} $a$, the \textit{argument of periapsis} $\omega$, the \textit{longitude of the ascending node} $\Omega$, and the orbital \textit{inclination} $i$. The orbit of a binary system, i.e., the position in the plane of the sky $(X(t),Y(t))$ at a given time $t$, can be calculated through the following steps:
\begin{enumerate}
  \item Determination of the so-called eccentric anomaly $E(t)$ at a certain epoch t by numerically solving Kepler´s equation:
  \begin{equation}
      E(t)-e\sin E(t)=2\pi(t-T)/P.
  \label{eccentric_anomaly}
  \end{equation}
  
  \item Calculation of the auxiliary normalized coordinates $(x(t),y(t))$:
  \begin{equation}
    \begin{split}
    x(t)&=\cos{E(t)}-e, \\
    y(t)&=\sqrt{1-e^2}\sin{E(t)}.
\end{split}
  \end{equation}
  
  \item Determination of the Thiele-Innes constants:
  \begin{equation}
    \begin{split}
      A&=a(\cos{\omega}\cos{\Omega}-\sin{\omega}\sin{\Omega}\cos{i}),\\
      B&=a(\cos{\omega}\sin{\Omega}+\sin{\omega}\cos{\Omega}\cos{i}),\\
      F&=a(-\sin{\omega}\cos{\Omega}-\cos{\omega}\sin{\Omega}\cos{i}),\\
      G&=a(-\sin{\omega}\sin{\Omega}+\cos{\omega}\cos{\Omega}\cos{i}).
    \end{split}
    \label{eq:thiele_innes}
  \end{equation}
  
  \item Calculation of position in the apparent orbit $(X(t),Y(t))$:
  \begin{equation}
    \begin{split}
        X(t)&=Ax(t)+Fy(t),\\
        Y(t)&=Bx(t)+Gy(t).
    \end{split}
    \label{eq:pos}
  \end{equation}
\end{enumerate}

To compute the RV of each component of the binary system $(V_{1}(t),V_{2}(t))$ for primary and secondary respectively), it is necessary to incorporate additional parameters: the parallax $\varpi$, the \textit{mass ratio} of the individual components $q=m_2/m_1$ and the \textit{velocity of the center of mass} $V_{0}$. Thereby, the calculation of RV of each component of the system involves the following steps:
\begin{enumerate}
  \item Determination of the true anomaly $\nu(t)$ at a specific time $t$ using the eccentric anomaly $E(t)$ determined in Equation~(\ref{eccentric_anomaly}):
  \begin{equation}
    \tan{\dfrac{\nu(t)}{2}}=\sqrt{\dfrac{1+e}{1-e}}\tan{\dfrac{E(t)}{2}}.
  \label{eq:05}
  \end{equation}

  \item Calculation of the RV of the system´s individual components $(V_{1}(t),V_{2}(t))$:
  \begin{equation}
        V_{1}(t)=V_{0}+\dfrac{2\pi a_1 \sin{i}}{P\sqrt{1-e^2}}[\cos(\omega + \nu(t))+e\cos(\omega)],
  \label{eq:V_p}
  \end{equation}
  \begin{equation}
    V_{2}(t)=V_{0}-\dfrac{2\pi a_2 \sin{i}}{P\sqrt{1-e^2}}[\cos(\omega + \nu(t))+e\cos(\omega)],
    \label{eq:V_c}
  \end{equation}
  where $a_1=a''/\varpi\cdot q/(1+q)$, $a_2=a''/\varpi\cdot 1/(1+q)$ and $a''$ the semi-major axis in seconds of arc.
\end{enumerate}
Note that the determination of the true anomaly $\nu(t)$ in Equation~(\ref{eq:05}) presents no ambiguity, because this parameter has the same sign as the eccentric anomaly $E(t)$. Furthermore, the expression for the RV (\ref{eq:V_p}) and (\ref{eq:V_c}) contain the orbital parallax $\varpi$ explicitly, with the aim of exploding the full interdependence relations of the orbital parameters, avoiding to condense some of the parameters in those expressions as an independent parameter on the amplitude of the RV curve $K_1=(2\pi a_1\sin i)/(P\sqrt{1-e^2})$ in Equation~(\ref{eq:V_p}) and $K_2=(2\pi a_1\sin i)/(P\sqrt{1-e^2})$ in Equation~(\ref{eq:V_c}), as discussed in \cite{mendez2017orbits}.

If RV observations of each component ($V_1(t)$ and $V_2(t)$) are available ($SB2$ hereinafter), the combined model that describes the positional and RV observations is characterized by the vector of orbital parameters $\theta _{SB2}=\{P,T,e,a,\omega,\Omega,i,V_0,\varpi,q\}$. However, if the RV observations of only one component are available ($SB1$ case), the parameters $q$ and $\varpi$ cannot be simultaneously determined. In this case, the parameters $q$ and $\varpi$ can be condensed into the auxiliary parameter $f/\varpi$ \citep{videla2022bayesian}, where $f=q/(1+q)$ is the so called fractional-mass of the system. Thereby, the vector of orbital parameters that describes the positional and RV observations of single-line binaries with a visual orbit is $\theta_{SB1}=\{P,T,e,a,\omega,\Omega,i,V_0,f/\varpi\}$.

\subsection{Visual-triple hierarchical stellar system}

The hierarchical system approximation successively decomposes the system into two subgroups of stars whose dynamics follow the solution of the two-body problem. This results in a hierarchical structure of binary subsystems. As a consequence of the little interaction imposed by the hierarchical structure, the dynamics of each component follows an approximately stable Keplerian orbit around the center of mass of the system.

The simpler hierarchical system is composed of three stars or components. Two different hierarchical structures composed of two binary systems are possible: An outer object orbiting an inner binary system and an outer binary system orbiting an inner object.

Considering the first hierarchical structure: let $A_a$ and $A_b$ be the components of the inner binary system ($A_aA_b$). The motion of the third component is described by a Keplerian orbit around the center of mass of the inner system $A$, forming the outer binary system $AB$. 

According to the 2-body problem approximation, the force experimented by the companion object $A_b$ relative to the primary
object $A_a$ is given by:
\begin{equation}
    \ddot{\vec{r}}_{A_aA_b}=\ddot{\vec{r}}_{A_b}-\ddot{\vec{r}}_{A_a}=-\frac{G(m_{A_a}+m_{A_b})}{r_{A_aA_b}^2}\hat{r}_{A_aA_b}.
    \label{eq:r_AaAb}
\end{equation}

In the same manner, the force experimented by the relative position of the external object $B$ relative to the center of mass of the binary system $A_aA_b$ is given by:
\begin{equation}
    \ddot{\vec{r}}_{AB}=\ddot{\vec{r}}_{B}-\ddot{\vec{r}}_{A}=-\frac{G(m_{A}+m_{B})}{r_{AB}^2}\hat{r}_{AB}.
    \label{eq:r_AB}
\end{equation}

As explained in Appendix~\ref{App:binaries}, the solution of Equations~(\ref{eq:r_AaAb}) and (\ref{eq:r_AB}) are Keplerian elliptical orbits characterized by the vector of orbital parameters $\{P_{A_aA_b},T_{A_aA_b},e_{A_aA_b},a_{A_aA_b},\omega_{A_aA_b},\Omega_{A_aA_b},i_{A_aA_b}\}$ and $\{P_{AB},T_{AB},e_{AB},a_{AB},\omega_{AB},\Omega_{AB},i_{AB}\}$, respectively. The procedure to determine the position in those orbits at a certain time $t$ follows the ephemerides formulae described in Equation~(\ref{eq:pos}).

The observations of the inner and outer binary systems are generally measured relative to the brighter and most massive component of the system $A_a$ (principal component). Hence, as the solution of the inner system $\vec{r}_{A_aA_b}$ is already relative to $A_a$, it is necessary to characterize the position vector $\vec{r}_{A_aB}$ to describe the observations of the outer binary system.

The vector $\vec{r}_{A_aB}$ can be expressed as:
\begin{equation}
    \vec{r}_{A_aB}=\vec{r}_{A_aA}+\vec{r}_{AB}.
    \label{eq:r_AaB_desc}
\end{equation}
Decomposing the vector $\vec{r}_{A_aA}$ into its components and noting that $\vec{r}_A$ is the position of the center of mass of the inner binary system, we have that:
\begin{equation}
\begin{split}
    \vec{r}_{A_aA}&=\vec{r}_{A}-\vec{r}_{A_a}\\
    &=\frac{m_{A_a}\vec{r}_{A_a}+m_{A_b}\vec{r}_{A_b}}{m_{A_a}+m_{A_b}}-\vec{r}_{A_a}\\
    &=(\vec{r}_{A_b}-\vec{r}_{A_a})\frac{m_{A_b}}{m_{A_a}+m_{A_b}}\\
    &=\vec{r}_{A_aA_b}\left(\frac{q_{A_aA_b}}{1+q_{A_aA_b}}\right),
    \label{eq:r_Aa}
\end{split}
\end{equation}
with $q_{A_aA_b}=m_{A_b}/m_{A_a}$ denoting the mass ratio of the inner system. By replacing Equation~(\ref{eq:r_Aa}) into Equation~(\ref{eq:r_AaB_desc}), the position of the outer object relative to the system principal component becomes:
\begin{equation}
    \vec{r}_{A_aB}=\vec{r}_{AB}+\vec{r}_{A_aA_b}\left(\frac{q_{A_aA_b}}{1+q_{A_aA_b}}\right),
     \label{Sec1:pos_AaB}
\end{equation}
and therefore, the observations of the outer component relative to the principal component of the inner system is characterized by the vector of orbital parameters $\{P_{A_aA_b},T_{A_aA_b},e_{A_aA_b},a_{A_aA_b},\omega_{A_aA_b},\Omega_{A_aA_b},\\i_{A_aA_b}\}\cup\{q_{A_aA_b}\}\cup\{P_{AB},T_{AB},e_{AB},a_{AB},\omega_{AB},\Omega_{AB},i_{AB}\}$. It is important to note that while the positional vectors $\vec{r}_{A_aA_b}$ and $\vec{r}_{AB}$ follow a Keplerian orbit (since they are the solution of the 2-body problem of the binary systems $A_aA_b$ and $AB$, respectively), the positional vector $\vec{r}_{A_aB}$ does not, being a non-closed orbit. A relevant derived orbital parameter that characterizes the interaction between the inner and outer system, denoted as the mutual inclination $\Phi$, is defined as follows:
\begin{equation}
    \cos(\Phi) = \cos(i_{A_aA_b})\cdot\cos(i_{AB}) + \sin(i_{A_aA_b})\cdot\sin(i_{AB})\cdot \cos(\Omega_{AB} - \Omega_{A_aA_b}). \label{eq:mutual}
\end{equation}

Now considering the alternative hierarchical structure: let $A$ be the principal component and $B_a,B_b$ be the components of the outer binary system. The motion of the center of mass of the outer system $B$  is described by a Keplerian orbit around the inner component $A$, forming the inner binary system $AB$. 

Then, an analogous procedure can be developed to express the outer binary system observations in terms of positional vectors that are the solution of the 2-body problem characterized by its respective orbital parameters. Therefore, the positional vector that describes the outer system observations $\vec{r}_{AB_a}$ can be expressed as:
\begin{equation}
\begin{split}
    \vec{r}_{AB_a}&=\vec{r}_{AB}+\vec{r}_{BB_a}\\&=\vec{r}_{AB}-\vec{r}_{B_aB}\\
    &=\vec{r}_{AB}-\vec{r}_{B_aB_b}\left(\frac{q_{B_aB_b}}{1+q_{B_aB_b}}\right),
    \label{Sec1:pos_BaA}
\end{split}
\end{equation}
where the last equality follows the result shown in Equation~(\ref{eq:r_Aa}).

\section{Complementary results}
\label{App:complementary_results}

\begin{table}[h]
\centering
\caption{MAP estimates and 95\% HDPIs from the marginal posterior distributions for the orbital elements and derived physical parameters of the binary HIP 89000 after incorporating virtual observations $d_{\tau_i}\sim\mathcal{N}(f_{\text{kep}}(\theta_{map},\tau_i),\sigma^2),i\in\{1,2,3,4,5\}$, with $\sigma^2$ fixed to the minimum variance of the available system's observations $\mathcal{D}$.}
\begin{tabular}{lllllll}
\hline\hline
         $\theta$ &                            $SB2$ &                 $SB2+d_{\tau_1}$ &                 $SB2+d_{\tau_2}$ &                 $SB2+d_{\tau_3}$ &                 $SB2+d_{\tau_4}$ &                 $SB2+d_{\tau_5}$ \\
\hline
       $P$ $(yr)$ &          $0.546_{0.546}^{0.547}$ &          $0.546_{0.546}^{0.547}$ &          $0.546_{0.546}^{0.547}$ &          $0.546_{0.546}^{0.547}$ &          $0.546_{0.546}^{0.547}$ &          $0.546_{0.546}^{0.547}$ \\
       $T$ $(yr)$ & $1990.675_{1990.673}^{1990.677}$ & $1990.675_{1990.673}^{1990.677}$ & $1990.675_{1990.673}^{1990.677}$ & $1990.675_{1990.673}^{1990.677}$ & $1990.675_{1990.673}^{1990.677}$ & $1990.675_{1990.673}^{1990.677}$ \\
              $e$ &          $0.302_{0.297}^{0.306}$ &          $0.302_{0.297}^{0.306}$ &          $0.302_{0.297}^{0.307}$ &          $0.302_{0.298}^{0.306}$ &          $0.302_{0.297}^{0.306}$ &          $0.302_{0.297}^{0.306}$ \\
        $a$ $('')$ &          $0.021_{0.017}^{0.023}$ &          $0.021_{0.018}^{0.023}$ &          $0.021_{0.019}^{0.023}$ &          $0.021_{0.018}^{0.023}$ &          $0.021_{0.019}^{0.022}$ &          $0.021_{0.019}^{0.023}$ \\
   $\omega$ (\textdegree) &       $86.775_{85.739}^{87.441}$ &       $86.637_{85.793}^{87.473}$ &       $86.736_{85.776}^{87.460}$ &       $86.715_{85.878}^{87.491}$ &       $86.621_{85.756}^{87.450}$ &       $86.484_{85.806}^{87.517}$ \\
   $\Omega$ (\textdegree) &       $49.611_{45.878}^{54.205}$ &       $50.965_{45.904}^{54.065}$ &       $49.904_{45.660}^{54.080}$ &       $49.413_{46.258}^{53.735}$ &       $49.872_{46.138}^{54.147}$ &       $49.449_{46.025}^{54.198}$ \\
        $i$ (\textdegree) &    $141.080_{133.203}^{157.739}$ &    $140.322_{133.265}^{155.190}$ &    $141.324_{134.968}^{148.851}$ &    $141.579_{133.345}^{157.203}$ &    $142.721_{135.587}^{148.903}$ &    $141.400_{135.057}^{150.454}$ \\
 $V_{0}$ $(km/s)$ &    $-14.124_{-14.174}^{-14.091}$ &    $-14.130_{-14.173}^{-14.087}$ &    $-14.145_{-14.172}^{-14.086}$ &    $-14.119_{-14.174}^{-14.089}$ &    $-14.140_{-14.173}^{-14.088}$ &    $-14.137_{-14.174}^{-14.089}$ \\
 $\varpi$ $(mas)$ &       $26.212_{12.424}^{33.052}$ &       $26.846_{14.126}^{32.800}$ &       $25.870_{19.709}^{31.559}$ &       $25.658_{13.108}^{33.083}$ &       $24.698_{19.557}^{30.902}$ &       $25.494_{18.180}^{31.334}$ \\
$m_1$ $(M_\odot)$ &          $0.894_{0.482}^{3.423}$ &          $0.854_{0.518}^{2.706}$ &          $0.909_{0.592}^{1.530}$ &          $0.925_{0.510}^{3.268}$ &          $1.004_{0.610}^{1.529}$ &          $0.919_{0.595}^{1.754}$ \\
$m_2$ $(M_\odot)$ &          $0.855_{0.473}^{3.287}$ &          $0.817_{0.498}^{2.592}$ &          $0.870_{0.563}^{1.460}$ &          $0.885_{0.476}^{3.109}$ &          $0.960_{0.579}^{1.461}$ &          $0.878_{0.560}^{1.666}$ \\
              $q$ &          $0.956_{0.950}^{0.963}$ &          $0.956_{0.950}^{0.963}$ &          $0.956_{0.949}^{0.963}$ &          $0.957_{0.949}^{0.963}$ &          $0.956_{0.949}^{0.963}$ &          $0.956_{0.950}^{0.963}$ \\
\hline
\end{tabular}

\label{tab:YSC132AaAb_params}
\end{table}

\begin{table}[h]
\centering
\caption{MAP estimates and 95\% HDPIs from the marginal posterior distributions for the orbital elements and derived physical parameters of the binary HIP 99675 after incorporating virtual observations $d_{\tau_i}\sim\mathcal{N}(f_{\text{kep}}(\theta_{map},\tau_i),\sigma^2),i\in\{1,2,3,4,5\}$, with $\sigma^2$ fixed to the minimum variance of the available system's observations $\mathcal{D}$.}
\begin{tabular}{lllllll}
\hline\hline
         $\theta$ &                            $SB1$ &                 $SB1+d_{\tau_1}$ &                 $SB1+d_{\tau_2}$ &                 $SB1+d_{\tau_3}$ &                 $SB1+d_{\tau_4}$ &                 $SB1+d_{\tau_5}$ \\
\hline
       $P$ $(yr)$ &       $10.297_{10.265}^{10.332}$ &       $10.296_{10.266}^{10.333}$ &       $10.300_{10.263}^{10.332}$ &       $10.311_{10.264}^{10.332}$ &       $10.297_{10.266}^{10.333}$ &       $10.300_{10.265}^{10.333}$ \\
       $T$ $(yr)$ & $1950.528_{1950.372}^{1950.675}$ & $1950.540_{1950.374}^{1950.674}$ & $1950.526_{1950.370}^{1950.671}$ & $1950.485_{1950.366}^{1950.675}$ & $1950.540_{1950.376}^{1950.676}$ & $1950.518_{1950.370}^{1950.674}$ \\
              $e$ &          $0.202_{0.195}^{0.210}$ &          $0.202_{0.195}^{0.210}$ &          $0.203_{0.195}^{0.211}$ &          $0.203_{0.195}^{0.210}$ &          $0.202_{0.195}^{0.210}$ &          $0.202_{0.195}^{0.210}$ \\
        $a$ $('')$ &          $0.031_{0.025}^{0.036}$ &          $0.031_{0.030}^{0.032}$ &          $0.031_{0.028}^{0.034}$ &          $0.032_{0.030}^{0.033}$ &          $0.032_{0.027}^{0.036}$ &          $0.032_{0.029}^{0.033}$ \\
   $\omega$ (\textdegree) &    $198.678_{196.574}^{201.382}$ &    $199.033_{196.532}^{201.309}$ &    $199.001_{196.558}^{201.194}$ &    $199.141_{196.539}^{201.244}$ &    $199.217_{196.608}^{201.293}$ &    $198.653_{196.672}^{201.341}$ \\
   $\Omega$ (\textdegree) &    $305.271_{294.422}^{317.330}$ &    $304.496_{298.794}^{313.434}$ &    $304.182_{294.971}^{313.971}$ &    $305.464_{302.494}^{307.996}$ &    $305.521_{299.980}^{311.866}$ &    $307.406_{295.709}^{313.424}$ \\
        $i$ (\textdegree) &     $110.204_{79.770}^{132.995}$ &     $108.993_{97.224}^{127.468}$ &    $111.385_{102.014}^{121.306}$ &     $106.681_{83.162}^{131.474}$ &    $109.769_{107.044}^{114.951}$ &     $106.953_{98.631}^{126.627}$ \\
 $V_{0}$ $(km/s)$ &       $-6.390_{-6.470}^{-6.284}$ &       $-6.389_{-6.474}^{-6.291}$ &       $-6.377_{-6.467}^{-6.278}$ &       $-6.355_{-6.470}^{-6.286}$ &       $-6.381_{-6.472}^{-6.288}$ &       $-6.383_{-6.477}^{-6.289}$ \\
$f/\varpi$ $(pc)$ &    $161.221_{142.424}^{199.983}$ &    $161.508_{147.767}^{192.703}$ &    $162.614_{143.082}^{193.141}$ &    $158.391_{147.878}^{196.190}$ &    $160.074_{140.010}^{196.435}$ &    $158.118_{147.137}^{194.824}$ \\
\hline
\end{tabular}

\label{tab:HIP99675_params}
\end{table}

\begin{table}[h]
\centering
\caption{MAP estimates and 95\% HDPIs from the marginal posterior distributions for the orbital elements and derived physical parameters of the hierarchical triple system LHS 1070 after incorporating virtual observations $d_{\tau_i}\sim\mathcal{N}(f_{\text{kep}}(\theta_{map},\tau_i),\sigma^2),i\in\{1,2,3,4,5\}$, with $\sigma^2$ fixed to the minimum variance of the available system's observations $\mathcal{D}$.}
\begin{tabular}{lllllll}
\hline\hline
             $\theta$ &                    $AS_{triple}$ &         $AS_{triple}+d_{\tau_1}$ &         $AS_{triple}+d_{\tau_2}$ &         $AS_{triple}+d_{\tau_3}$ &         $AS_{triple}+d_{\tau_4}$ &         $AS_{triple}+d_{\tau_5}$ \\
\hline
    $P_{BaBb}$ $(yr)$ &       $17.282_{17.261}^{17.290}$ &       $17.280_{17.261}^{17.289}$ &       $17.270_{17.261}^{17.289}$ &       $17.274_{17.262}^{17.290}$ &       $17.274_{17.261}^{17.290}$ &       $17.276_{17.263}^{17.288}$ \\
    $T_{BaBb}$ $(yr)$ & $2005.081_{2004.754}^{2005.752}$ & $2005.180_{2004.749}^{2005.778}$ & $2005.174_{2004.725}^{2005.754}$ & $2005.271_{2004.731}^{2005.777}$ & $2005.130_{2004.738}^{2005.754}$ & $2005.085_{2004.750}^{2005.753}$ \\
           $e_{BaBb}$ &          $0.014_{0.013}^{0.016}$ &          $0.015_{0.013}^{0.016}$ &          $0.014_{0.013}^{0.016}$ &          $0.015_{0.013}^{0.016}$ &          $0.015_{0.013}^{0.016}$ &          $0.015_{0.013}^{0.016}$ \\
     $a_{BaBb}$ $('')$ &          $0.463_{0.462}^{0.464}$ &          $0.463_{0.462}^{0.464}$ &          $0.463_{0.462}^{0.464}$ &          $0.463_{0.462}^{0.464}$ &          $0.463_{0.462}^{0.464}$ &          $0.463_{0.462}^{0.464}$ \\
$\omega_{BaBb}$ (\textdegree) &    $174.313_{167.394}^{188.218}$ &    $176.356_{167.359}^{188.905}$ &    $176.142_{166.909}^{188.365}$ &    $178.222_{166.951}^{188.781}$ &    $175.308_{167.223}^{188.416}$ &    $174.283_{167.142}^{188.082}$ \\
$\Omega_{BaBb}$ (\textdegree) &       $14.608_{14.404}^{14.842}$ &       $14.612_{14.393}^{14.823}$ &       $14.695_{14.389}^{14.841}$ &       $14.626_{14.390}^{14.832}$ &       $14.572_{14.382}^{14.830}$ &       $14.666_{14.394}^{14.834}$ \\
     $i_{BaBb}$ (\textdegree) &       $62.962_{62.844}^{63.199}$ &       $62.975_{62.820}^{63.178}$ &       $62.964_{62.840}^{63.200}$ &       $63.024_{62.832}^{63.196}$ &       $63.063_{62.828}^{63.189}$ &       $63.001_{62.837}^{63.196}$ \\
           $q_{BaBb}$ &          $0.988_{0.979}^{1.000}$ &          $0.991_{0.979}^{1.000}$ &          $0.989_{0.979}^{1.000}$ &          $0.993_{0.978}^{1.000}$ &          $0.990_{0.979}^{1.000}$ &          $0.994_{0.979}^{1.000}$ \\
      $P_{AB}$ $(yr)$ &     $105.414_{88.241}^{124.480}$ &    $110.701_{101.827}^{118.228}$ &    $109.692_{106.627}^{112.803}$ &    $108.935_{108.388}^{109.430}$ &    $108.699_{108.560}^{108.907}$ &    $108.693_{108.592}^{108.790}$ \\
      $T_{AB}$ $(yr)$ & $2001.345_{1996.313}^{2003.964}$ & $2002.556_{2000.305}^{2003.864}$ & $2002.231_{2001.257}^{2003.109}$ & $2001.952_{2001.636}^{2002.292}$ & $2001.803_{2001.574}^{2002.204}$ & $2002.162_{2000.923}^{2003.167}$ \\
             $e_{AB}$ &          $0.166_{0.071}^{0.259}$ &          $0.191_{0.150}^{0.227}$ &          $0.187_{0.175}^{0.200}$ &          $0.183_{0.182}^{0.184}$ &          $0.183_{0.181}^{0.184}$ &          $0.181_{0.177}^{0.186}$ \\
       $a_{AB}$ $('')$ &          $1.886_{1.665}^{2.117}$ &          $1.947_{1.843}^{2.036}$ &          $1.935_{1.907}^{1.968}$ &          $1.927_{1.923}^{1.931}$ &          $1.926_{1.919}^{1.932}$ &          $1.921_{1.913}^{1.934}$ \\
  $\omega_{AB}$ (\textdegree) &    $349.818_{328.230}^{359.905}$ &    $354.614_{345.724}^{359.512}$ &    $353.399_{349.646}^{356.914}$ &    $352.330_{350.987}^{353.721}$ &    $351.754_{350.899}^{353.264}$ &    $353.111_{348.703}^{357.096}$ \\
  $\Omega_{AB}$ (\textdegree) &       $13.263_{12.381}^{14.166}$ &       $13.699_{12.789}^{14.249}$ &       $13.535_{13.131}^{13.921}$ &       $13.409_{13.281}^{13.561}$ &       $13.334_{13.180}^{13.572}$ &       $13.530_{12.868}^{14.017}$ \\
       $i_{AB}$ (\textdegree) &       $65.577_{63.908}^{66.793}$ &       $65.800_{65.277}^{66.258}$ &       $65.753_{65.615}^{65.949}$ &       $65.745_{65.586}^{65.909}$ &       $65.753_{65.605}^{65.898}$ &       $65.652_{65.442}^{65.979}$ \\
         $\Phi$ (\textdegree) &          $2.882_{1.646}^{3.864}$ &          $2.943_{2.599}^{3.325}$ &          $2.979_{2.684}^{3.200}$ &          $2.933_{2.694}^{3.178}$ &          $2.913_{2.737}^{3.186}$ &          $2.842_{2.446}^{3.332}$ \\
\hline
\end{tabular}

\label{tab:LHS1070_params}
\end{table}

\bibliography{sample631}{}
\bibliographystyle{aasjournal}

%% This command is needed to show the entire author+affiliation list when
%% the collaboration and author truncation commands are used.  It has to
%% go at the end of the manuscript.
%\allauthors

%% Include this line if you are using the \added, \replaced, \deleted
%% commands to see a summary list of all changes at the end of the article.
%\listofchanges

\end{document}